\documentclass[%
reprint, twocolumn, superscriptaddress,
showpacs,preprintnumbers,
nofootinbib,
 amsmath,amssymb,
 aps,
prl,
]{revtex4}
\usepackage{graphicx}
\usepackage{amssymb}
\usepackage{amsmath}
\usepackage{color}

\newcommand {\fabsq}[1] {\left| #1 \right|^2}
\newcommand {\fabs}[1] {\left| #1 \right|}
\def\ra{\rangle}
\def\la{\langle}

\newcommand{\beqa}{\begin{eqnarray}}
\newcommand{\eeqa}{\end{eqnarray}}
\newcommand{\beq}{\begin{equation}}
\newcommand{\eeq}{\end{equation}}

\begin{document}
\title{Asymmetric scattering by non-hermitian potentials}
\author{A. Ruschhaupt}
\affiliation{Department of Physics, University College Cork, Ireland}
\author{T. Dowdall}
\affiliation{Department of Physics, University College Cork, Ireland}
\author{M. A. Sim\'on}
\affiliation{Departamento de Qu\'{\i}mica F\'{\i}sica, UPV/EHU, Apdo.
644, 48080 Bilbao, Spain}
\author{J. G. Muga}
\affiliation{Departamento de Qu\'{\i}mica F\'{\i}sica, UPV/EHU, Apdo.
644, 48080 Bilbao, Spain}
\begin{abstract}
The scattering of quantum particles by non-hermitian (generally nonlocal)  potentials in one dimension
may result in asymmetric transmission and/or reflection from left and right incidence.
Eight generalized symmetries based on the discrete  Klein's four-group
(formed by parity, time reversal, their product, and unity) are used together with generalized unitarity relations to determine
the possible and/or forbidden scattering asymmetries.
Six basic device types are
identified when the scattering coefficients (squared moduli of scattering amplitudes)
adopt zero/one values, and transmission and/or reflection are asymmetric.
They can pictorically be described as a
one-way mirror, a one-way barrier (a Maxwell  pressure demon), one-way (transmission or reflection) filters,
a mirror with unidirectional transmission, and a transparent, one-way reflector. We design potentials
for these devices and also demonstrate that the  behavior of the scattering
coefficients can be extended to a broad range of incident momenta.
\end{abstract}
\pacs{03.65.Nk, 11.30.Er}
\maketitle
%
%
%
%
{\it Introduction.} The current interest to develop new quantum technologies is boosting applied and fundamental
research on quantum phenomena and on systems with potential applications in logic circuits, metrology, communications or sensors.
Robust basic devices performing elementary operations are needed to perform
complex tasks when combined in a circuit.

In this paper we investigate the properties of potentials with asymmetric
transmission or reflection for a quantum, spinless particle of mass $m$ satisfying a one-dimensional (1D) Schr\"odinger equation.
If we restrict the analysis  to  transmission and reflection coefficients (squared moduli
of the scattering complex amplitudes) being  either zero or one, a useful simplification for quantum logic operations,
there are  six types of asymmetric devices.
These devices cannot be constructed with Hermitian potentials. In fact for all
(four) device types with transmission asymmetries,
the potentials have to be also nonlocal.
Therefore, nonlocal potentials play a major role in this paper. They appear naturally when applying partitioning techniques
under similar conditions to the
ones leading to non-hermitian  potentials, namely, as effective interactions for a subsystem or component of the full wave-function,
even if the interactions for the large system are hermitian and local \cite{cp}.

Symmetries can be used, similarly to their standard application to determine allowed/forbidden
transitions in atomic physics,
to predict whether a certain potential may or may not lead to asymmetric
scattering. The concept of symmetry, however, must be generalized when dealing with non-hermitian potentials.



\begin{table*}
\centering
\scalebox{0.85}{
\begin{tabular}{|c|c|c|c|c|c|c|c|c|c|c|c|}
\hline
Code & Symmetry&  $\la x|V|y\ra=$ & $\la p|V|p'\ra=$ & $\la p|S|p'\ra=$ & $T^l=$ & $T^r=$ & $R^l=$& $R^r=$& from Eq. (S8)&$|T^l|=1,|T^r|=0$&$|R^l|=1,|R^r|=0$
\\
\hline
I & $1H=H1$ &   $\la x|V|y\ra$ & $\la p|V|p'\ra$ & $\la p|S|p'\ra$ & $T^l$ & $T^r$ & $R^l$ & $R^r$ & & $P$ & $P$
\\
II & $1H=H^\dagger 1$ &  $\la y|V|x\ra^*$ & $\la p'|V|p\ra^*$ &$\la p|\widehat{S}|p'\ra$ & $\widehat{T}^l$& $\widehat{T}^r$ & $\widehat{R}^l$ & $\widehat{R}^r$
& $|T^l|=|T^r|$, $|R^l|=|R^r|$&No&No
\\
III & $\Pi H=H\Pi$ &  $\la -x|V|-y\ra$ & $\la -p|V|-p'\ra$ &$\la -p|S|-p'\ra$ & $T^r$ & $T^l$ & $R^r$ & $R^l$&&No&No
\\
IV & $\Pi H=H^\dagger \Pi$ &  $\la -y|V|-x\ra^*$ & $\la -p'|V|-p\ra^*$ & $\la -p|\widehat{S}|-p'\ra$ & $\widehat{T}^r$ & $\widehat{T}^l$ & $\widehat{R}^r$ & $\widehat{R}^l$&&$P$, $R^rR^{l*}=1$&$P$, $T^r{T^l}^*=1$
\\
V & $\Theta H=H\Theta$ &  $\la x|V|y\ra^*$& $\la -p|V|-p'\ra^*$ & $\la -p'|\widehat{S}|-p\ra$ & $\widehat{T}^r$ & $\widehat{T}^l$ & $\widehat{R}^l$& $\widehat{R}^r$
&$|R^l|=|R^r|$&$P$, $|R^{r,l}|=1$&No
\\
VI & $\Theta H=H^\dagger\Theta$ &  $\la y|V|x\ra$& $\la -p'|V|-p\ra$ & $\la -p'|S|-p\ra$ & $T^r$& $T^l$ & $R^l$& $R^r$&&No&$P$
\\
VII & $\Theta\Pi H=H\Theta \Pi$ &  $\la -x|V|-y\ra^*$ & $\la p|V|p'\ra^*$ & $\la p'|\widehat{S}|p\ra$ &$\widehat{T}^l$& $\widehat{T}^r$ & $\widehat{R}^r$& $\widehat{R}^l$&$|T^l|=|T^r|$&No&$P$, $|T^{r,l}|=1$
\\
VIII& $\Theta\Pi H=H^\dagger \Theta \Pi$ &  $\la -y|V|-x\ra$ & $\la p'|V|p\ra$ & $\la p'|S|p\ra$ & $T^l$ & $T^r$ & $R^r$ & $R^l$&&$P$&No
\\
\hline
\end{tabular}
}
\caption{Symmetries of the potential classified in terms of the commutativity or pseudo-hermiticity of $H$ with the elements of
Klein's 4-group  $\{1,\Pi,\theta,\Pi\theta\}$ (second column). The first column sets a simplifying roman-number code for each symmetry.
The relations among potential matrix elements are given in coordinate and momentum representations in the third and fourth columns.
The fifth column gives the relations they imply in the matrix elements of $S$ and/or $\widehat{S}$ matrices ($S$ is for scattering by $H$
and $\widehat{S}$ for scattering by $H^\dagger$). From them  the next four columns set the relations implied on scattering amplitudes.
Together with generalized unitarity relations (S8) they also imply relations for the moduli (tenth column), and phases (not shown). The last two columns indicate the possibility to achieve perfect asymmetric transmission or reflection:  ``${P}$" means possible (but not necessary),
``No'' means impossible.  In some cases ``$P$" is accompanied by a condition that must be satisfied.
\label{table2}}
\end{table*}


{\it{Generalized symmetries.}}
The detailed technical and formal background for the following can be found in
a previous review on 1D scattering by complex potentials \cite{cp}, a companion to this article
for those readers willing to reproduce the calculations in detail.
The Supplemental Material (Sec. I)  provides also a minimal kit of scattering theory formulae that may be
read first to set basic concepts and notation.
The notation is essentially as in \cite{cp}, but it proves convenient to use
for the potential matrix (or kernel function) in coordinate representation
two different forms, namely $\la x|V|y\ra=V(x,y)$. ``Local'' potentials are those
for which $V(x,y)=V(x)\delta(x-y)$.

%
For hermitian  Hamiltonians, symmetries are represented by the commutation of
a symmetry operator with the Hamiltonian.
In scattering theory, symmetry plays an important role  as it implies relations among
the S-matrix elements beyond those implied by its unitarity, see e.g. \cite{Taylor} and, for scattering in one dimension,
Sec. 2.6 in \cite{cp}.

Symmetries are also useful for  non-hermitian Hamiltonians, but the mathematical and conceptual
framework must be generalized. We consider that a unitary or antiunitary
%
operator $A$ represents a symmetry of $H$ if it satisfies
at least one of these relations,
\beqa
AH&=&HA,
\label{symm}
\\
AH&=&H^\dagger A.
\label{pseudo}
\eeqa
%
When the symmetry in Eq. (\ref{pseudo}) holds we say that $H$ is $A$-pseudohermitian \cite{mostareview}.
Parity-pseudohermiticity has played an important role as being equivalent to space-time reflection (PT) symmetry for {\it local} potentials
 \cite{mostareview,znojil}. A large set of these equivalences will be discussed below.
A relation of the form (\ref{pseudo}) has been also used with differential operators  to get real spectra beyond
PT-symmetry for local potentials  \cite{NY2016a,NY2016b}.

Here we consider
$A$ to be a member of the
Klein 4-group $K_4=\{1,\Pi, \Theta, \Pi\Theta\}$ formed by unity, the parity operator $\Pi$, the antiunitary time-reversal operator $\Theta$, and their product
$\Pi\Theta$. This is a discrete, abelian group.
We also assume that the  Hamiltonian is  of the form $H=H_0+V$, with $H_0$, the kinetic energy operator of the particle, hermitian and
satisfying $[H_0,A]=0$ for all members of the group, whereas the potential $V$ may be non-local in position representation.
The  motivation to use Klein's group is that the eight relations implied by Eqs. (\ref{symm}) and (\ref{pseudo}) generate all
possible symmetries of a non-local potential due to the identity, complex conjugation, transposition, and sign inversion,
both in coordinate or momentum representation, see Table \ref{table2}, where each symmetry has been labeled by a roman number.
Interesting enough, in this classification hermiticity (II) may be regarded as $1$-pseudohermiticity.

Examples on how to find the relations in the fifth column of Table \ref{table2} of $S-$ and $\widehat{S}$-matrix elements (for scattering by $H$ and $H^\dagger$ respectively)
are provided in
ref. \cite{cp}, where the symmetry types III, VI, and VII where worked out. Similar manipulations, making use of the action of unitary or antiunitary operators
of Klein's group on M\"oller operators, help to
complete the  table.
From this fifth column, equivalences among the amplitudes for left and right incidence
for scattering by $H$, ($T^{l,r}, R^{l,r}$) or $H^\dagger$  ($\widehat{T}^{l,r}, \widehat{R}^{l,r}$),
are deduced, see the Supplemental Material  and the four columns for
$T^{l,r}$, and $R^{l,r}$.
Together with the generalized unitarity relations, see Eq. (S8),
these relations between the amplitudes
 imply further consequences on the amplitudes' moduli (tenth column of Table \ref{table2}) and phases (not shown).
 The final two columns use the previous results to determine if perfect asymmetry is possible for transmission or reflection.
 This makes evident that hermiticity (II) and parity (III) preclude, independently, any asymmetry in the scattering coefficients;
 PT-symmetry (VII) or  $\Theta$-pseudohermiticity
 (VI) forbid transmission asymmetry, whereas time-reversal symmetry (i.e., a real potential in coordinate space)
 (V) or  PT-pseudohermiticity (VIII) forbid reflection asymmetry.
 Note that all local potentials  satisfy automatically
 $\Theta$-pseudohermiticity (VI). Of course asymmetric effects forbidden by a certain symmetry in the linear (Schr\"odinger)
 regime considered in this paper might not be forbidden in a non-linear regime \cite{Xu}, which goes beyond our present scope.
 %
 %
 %
 %

 The occurrence of one particular symmetry in the potential (conventionally  ``first symmetry'')
 does not exclude a second symmetry to be satisfied as well.
 When a double symmetry holds, excluding the identity,  the ``first'' symmetry  implies the equivalence of the second symmetry with a third symmetry.
 We have already mentioned that $\Pi$-pseudohermiticity (IV) is equivalent to $PT$-symmetry (VII) for local potentials.
 Being local is just one particular way to satisfy symmetry VI, namely $\Theta$-pseudohermiticity. The reader may verify with the aid of
 the third column for $\la x|V|y\ra$  in Table \ref{table2}, that indeed, if symmetry VI is satisfied (first symmetry), symmetry IV has the same effect as symmetry VII.
 They become equivalent. Other well known example is the fact that for a local potential (symmetry VI is satisfied), a real potential in coordinate space  is necessarily hermitian,
 so symmetries V and II become equivalent.
 These are just particular cases of the full set of equivalences given in Table \ref{tablee}.
%
%
%
%
%
%
%
%

Combining the information of the last two-columns in Table \ref{table2} with the additional condition that all scattering coefficients
be 0/1 we elaborate Table \ref{table1}, which provides names for the six possible types of devices, a convenient letter code
that summarizes the effect of left/right incidence, and the symmetries
that do not allow the implementation of each device type.
The complementary Table \ref{table3} in the Supplemental Material gives instead the symmetries that allow, but do not necessarily imply,
a given device type.
The denominations in Table \ref{table1} are intended as short and meaningful, and do not necessarily coincide with
some extended terminology, in part because the range of possibilities is broader here than those customarily considered, and because we
use a 1 or 0 condition for the moduli.
For example, a device with reflection asymmetry with $T^r=T^l$ would in our case be a particular
``transparent, one-way reflector'', as full transmission occurs from both sides.
This effect has however become popularized as ``unidirectional invisibility'' \cite{unidirinv,Yin,optics}.
A debate on terminology is not our main concern here, and the use of a code system
as the one proposed will be instrumental in avoiding misunderstandings.

\begin{table*}
\centering
\scalebox{1.0}{
\begin{tabular}{|c|c|c|c|c|c|c|}
\hline
II & III& IV& V& VI & VII &VIII
\\
\hline
III=IV & II=IV & II=III & II=VI &II=V&II=VIII&II=VII
\\
V=VI&V=VII&V=VIII&III=VII&III=VIII&III=V&III=VI
\\
VII=VIII&VI=VIII&VI=VII&IV=VIII&IV=VII&IV=VI&IV=V
\\
\hline
\end{tabular}}
\caption{Equivalences among symmetries for the potential elements.
Given the symmetry of the upper row, the table provides the equivalent symmetries.
For example, if II is satisfied, then III=IV holds. In words, if the potential is hermitian,  parity symmetry amounts to
parity pseudohermiticity. In terms of the matrix elements of the potential, if  $\la x|V|y\ra=\la y|V|x\ra^*$ {\it and also}
$\la x|V|y\ra=\la -x|V|-y\ra$, $\forall (x,y)$, then $\la x|V|y\ra=\la -y|V|-x\ra^*$ holds as well. One may proceed similarly for all other relations.
The commutation with the identity (I) is excluded as this symmetry is satisfied by all potentials.
 \label{tablee}}
\end{table*}

\begin{table*}
\centering
\scalebox{1.0}{
\begin{tabular}{|c|c|c|c|c|c|c|}
\hline
Device type & Left incidence& Right incidence&Code& Forbidden by & Example
\\
\hline
One-way mirror&transmits and reflects&absorbs&$\cal{TR/A}$&II, III, IV, V, VI,  VII, VIII& Fig. \ref{fig_device_TR_A}
\\
One-way barrier&transmits&reflects&$\cal{T/R}$&II, III, IV, V, VI, VII, VIII&Fig. \ref{fig_device_T_R}
\\
One-way T-filter&transmits&absorbs&$\cal{T/A}$&II, III, IV, V, VI, VII&Fig. \ref{fig_device_T_A}
\\
Mirror\&1-way transmitter&transmits and reflects&reflects&$\cal{TR/R}$&II, III, VI, VII&Fig. \ref{fig_device_TR_R}
\\
One-way R-filter&reflects&absorbs&$\cal{R/A}$&II, III, IV, V, VII, VIII&[S1]
\\
Transparent 1-way reflector&transmits and reflects&transmits&$\cal{TR/T}$& II, III, V, VIII
& Figs. \ref{fig_reflector}, \ref{fig_nonlocal_pt}
\\
\hline
\end{tabular}}
\caption{Device types for  transmission and/or reflection asymmetry, restricted to (1/0) moduli for the scattering amplitudes.
The code summarizes the effect of left and right incidence, separated by a  slash $/$.  ${\cal T}$ or ${\cal R}$ on one side of the slash indicate a unit
transmission or reflection coefficient
for  incidence from that side, whereas the absence of one or the other letter corresponds to zero coefficients.
An ${\cal A}$ denotes ``full absorption'', i.e., both moduli of reflection and transmission amplitudes are zero for incidence from one side.
For example,  $\cal{TR/A}$ means unit modulus transmission
and reflection from the left and total absorption from the right.
The fifth column indicates the symmetries in Table \ref{table2} that forbid the device.\label{table1}}
\end{table*}

{\it{Designing potentials for asymmetric devices.\label{examples}}}
We will show  how to design non-local potentials
leading to the asymmetric devices.
For simplicity we look for  non-local potentials $V(x,y)$ with local support
that vanish  for $|x| >d$ and $|y| >d$.

Inverse scattering proceeds similarly to \cite{prl98},
by imposing an ansatz
for the wavefunctions and the potential
in the stationary Schr\"odinger equation
\begin{eqnarray}
\frac{\hbar^2k^2}{2m} \psi (x) = - \frac{\hbar^2}{2m} \frac{d^2}{dx^2} \psi (x)
+\!\!\int_{-d}^d \!dy V(x, y) \psi(y).
\label{Schroedinger}
\end{eqnarray}
The free parameters are fixed making use of the boundary conditions.
The expected form of the wavefunction incident from the left is
$\psi_l(x) = e^{i k x} + R^l e^{-i k x}$ for $x < -d$ and $\psi_l (x) = T^l e^{i k x}$ for $x > d$,
where  $k=p/\hbar$.
The wavefunction incident from the right is instead
$\psi_r(x) = e^{-ikx} T^r$ for $x < -d$ and $\psi_r (x) = e^{-i k x} + R^r e^{i k x}$ for $x > d$.

Our strategy is to assume  polynomial forms for the two wavefunctions in the interval $|x| < d$,
$\psi_l (x) = \sum_{j=0}^5 c_{l,j} x^j$ and $\psi_r (x) = \sum_{j=0}^5 c_{r,j} x^j$, and also a
polynomial ansatz of finite degree for the potential $V(x,y) = \sum_i \sum_j v_{ij} x^i y^j$.
Inserting these ansatzes in Eq. (\ref{Schroedinger}) and from the conditions that $\psi_{l,r}$
and their derivatives must be continuous, all coefficients $c_{l,j}\,,c_{r,j}$ and $v_{ij}$ can be determined.
Symmetry properties of the potential can also be imposed via additional conditions on
the potential coefficients $v_{ij}$.
In the  Supplemental Material, Sec. II,  we use this strategy to implement  potentials for
different devices in Table \ref{table1} such that for a chosen $k=k_0$ the imposed
boundary conditions (scattering amplitudes) are fulfilled exactly.
They are also satisfied approximately in a neighborhood of $k_0$.
%
%
%
%
%

%
\begin{figure}[hb]
\begin{center}
(a) \includegraphics[width = 0.6\columnwidth]{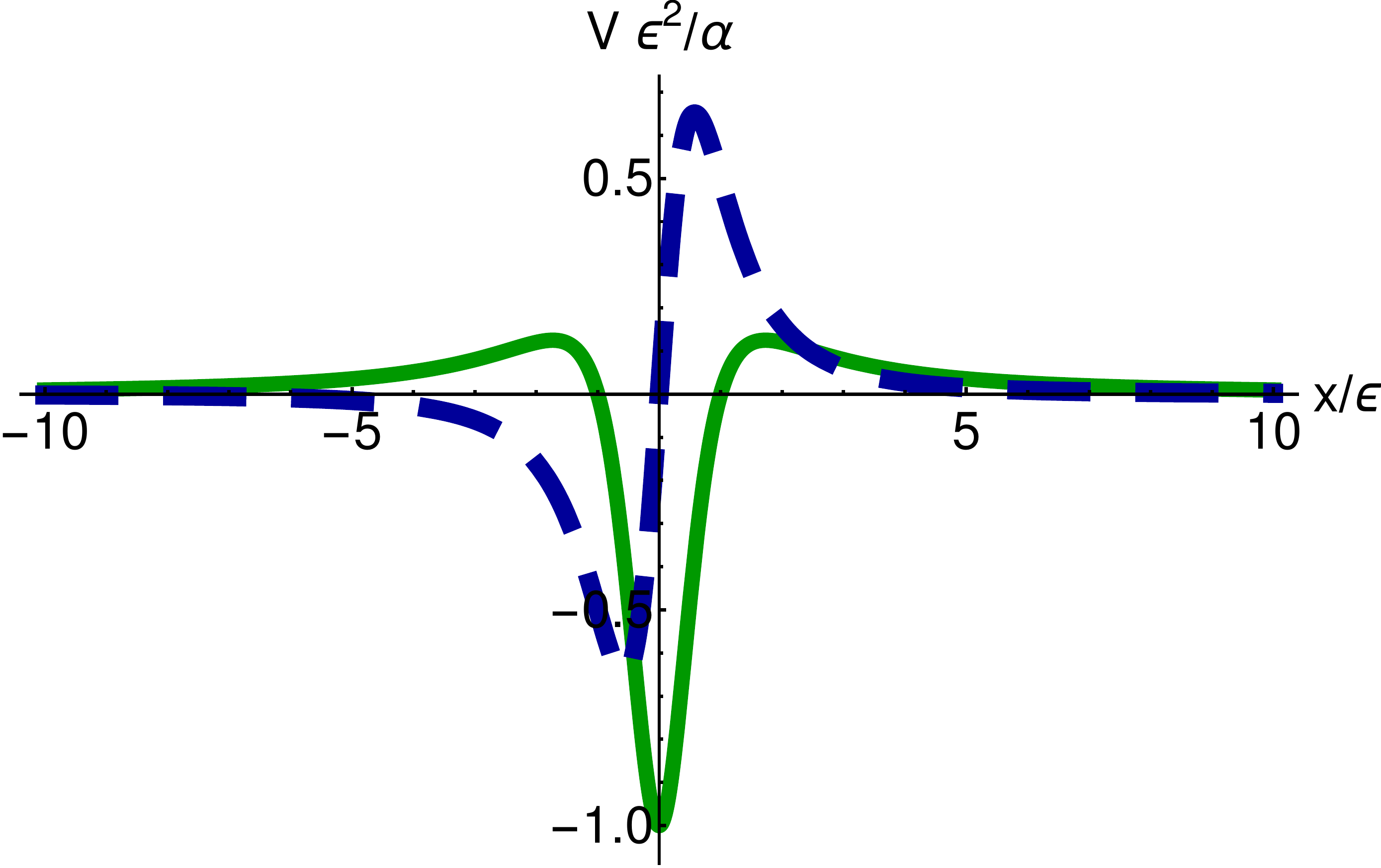}\\
(b) \includegraphics[width = 0.6\columnwidth]{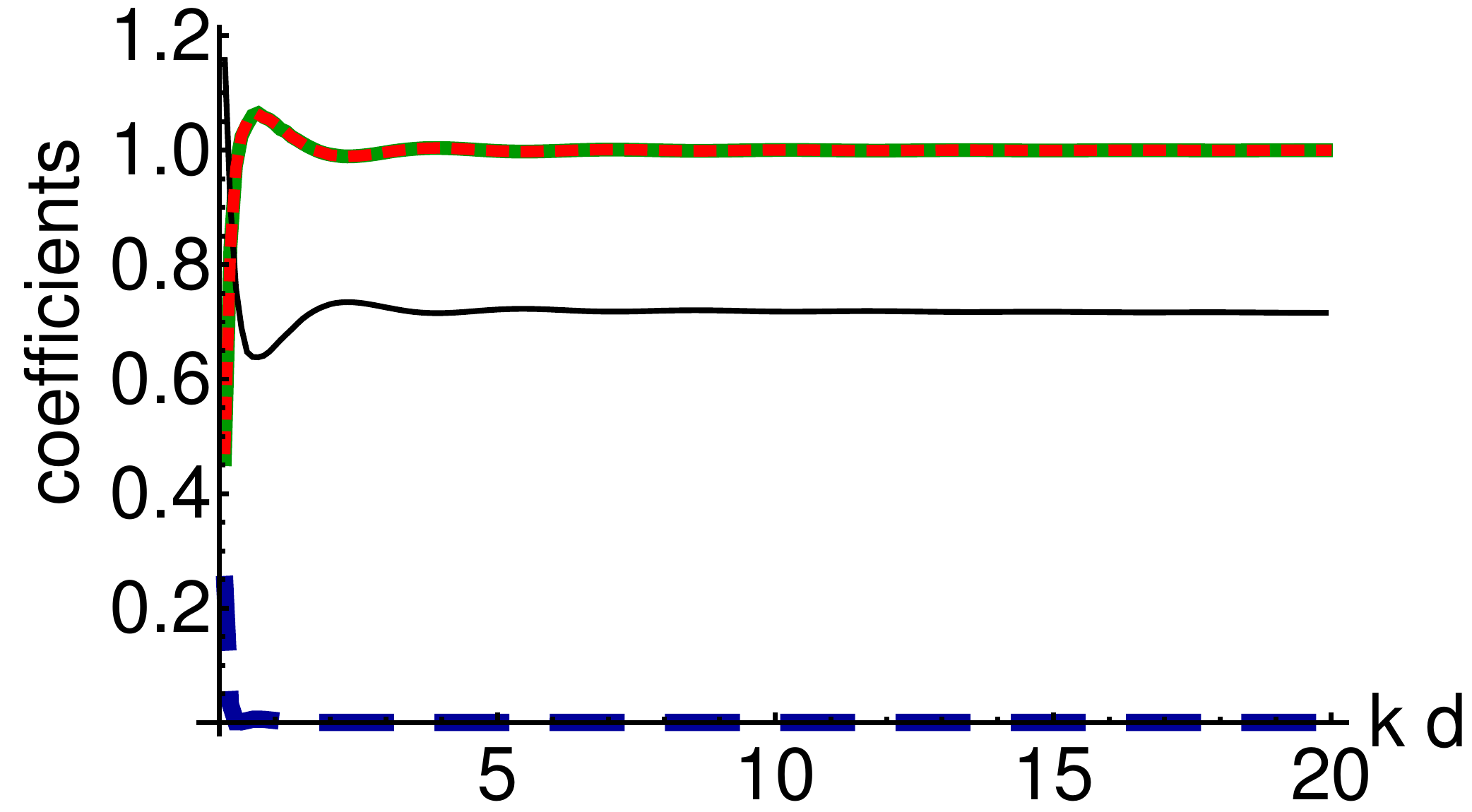}\\
(c) \includegraphics[width = 0.6\columnwidth]{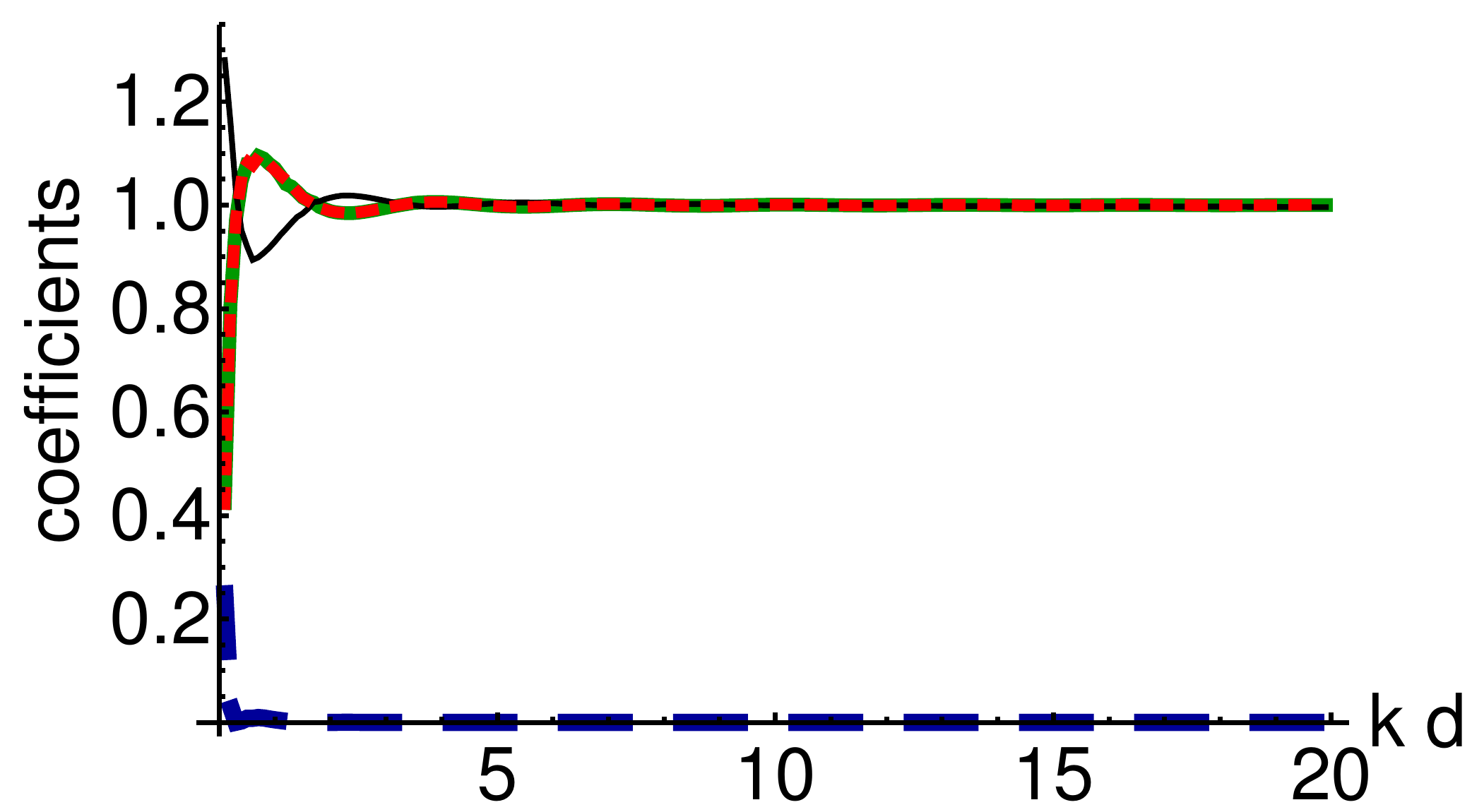}
\end{center}
\caption{\label{fig_reflector}(Color online) Transparent 1-way reflector with a local PT potential:
(a) Approximation of the potential Eq. (\ref{num}), real part (green solid line), imaginary part (blue dashed line).
(b,c) Transmission and reflection coefficients versus momentum $k d$;
left incidence: $\fabsq{R^l}$ (black, solid line), $\fabsq{T^l}$ (green, solid line);
right incidence: $\fabsq{R^r}$ (blue, tick, dashed line), $\fabsq{T^r}$ (red, dotted line, coincides with green, solid line).
$\epsilon/d = 10^{-4}$.
(b) $\alpha= 1.0 \hbar^2/(4\pi m)$ (c) $\alpha = 1.225 \hbar^2/(4\pi m)$
(the black, solid line coincides here mostly with the red, dotted and green, solid lines).
\label{fig_TR_T_local}}
\end{figure}
%

{\it{Extending the asymmetry to a broad incident-momentum domain\label{ext}}.}
The inversion technique just described may be generalized
to extend the range of incident momenta for which the potential works by imposing additional
conditions and increasing correspondingly the number of parameters in the wavefunction ansatz,
for example we may impose that the derivatives of the  amplitudes,  in one or more orders,  vanish at $k_0$,
or  0/1 values for the coefficients not only at  $k_0$ but at a series of grid points $k_1$, $k_2$, ... $k_N$,
as in \cite{BMM94,PM98,prl98,cp}.

Here we put forward instead a method that provides a very broad working-window domain
making use of the Born approximation.
Specifically we apply the approach for a transparent one-way reflector ${\cal{TR/T}}$.
The goal is now to find a local PT-symmetric potential such that asymmetric reflection results,
$T^l = T^r = 1, R^r = 0, |R^l|=1$ for a broad range of incident momenta. A similar goal
was pursued in \cite{Longhi2014} making use of a supersymmetric transformation,
without imposing $|R^l|=1$.

In the Born approximation and for a local potential $V(x)$, the reflection amplitudes take the simple form
\beqa
R^l=-\frac{2\pi i m}{p}\la -p|V|p\ra,
\;
R^r=-\frac{2\pi i m}{p}\la p|V|-p\ra.
\eeqa
Defining the Fourier transform
\begin{eqnarray}
\widetilde V (k) = \frac{1}{\sqrt{2\pi}} \int_{-\infty}^\infty dx \, V(x) e^{-i k x}
\end{eqnarray}
we get for $k=p/\hbar>0$:
\beqa
R^l=-\frac{\sqrt{2\pi} i m}{k \hbar^2} \widetilde V (-2k),
\;
R^r=-\frac{\sqrt{2\pi} i m}{k\hbar^2} \widetilde V (2 k).
\eeqa
Assuming that the potential is local and PT-symmetrical, we calculate the transition coefficient
from them using generalized unitarity as
$|T|^2=1-{R^r}^*R^l$.

To build a ${\cal{TR/T}}$ device we demand:
$\widetilde V(k) = \sqrt{2\pi} \alpha k$ ($k < 0$) and $\widetilde V(k) = 0$ ($k \ge 0$).
By inverse Fourier transformation, this implies
\begin{eqnarray}
V(x) &=&
-\alpha \frac{\partial}{\partial x} \lim_{\epsilon\to 0} \frac{1}{x - i \epsilon}
= \alpha \lim_{\epsilon\to 0} \frac{1}{(x - i \epsilon)^2}
\nonumber\\
&=& \alpha \lim_{\epsilon\to 0} \left[\frac{x^2 - \epsilon^2}{(x^2 + \epsilon^2)^2} + i
\frac{2 x\epsilon}{(x^2 + \epsilon^2)^2}\right],
\label{num}
\end{eqnarray}
which is indeed a local, $PT$-symmetric potential for $\alpha$ real.
$\alpha$ is directly related to the reflection coefficient, within the Born approximation,
$R^l = 4 \pi i m \alpha/\hbar^2$. As the Born approximation may differ from exact results
we shall keep $\alpha$ as an adjustable parameter
in the following.

In a possible physical implementation, the potential in Eq. (\ref{num}) will
be approximated by keeping a small finite $\epsilon>0$, see Fig. \ref{fig_TR_T_local} (a).
Then, its Fourier transform is
$\widetilde V(k) = \sqrt{2\pi} \alpha k e^{\epsilon k}$ ($k < 0$) and $\widetilde V(k)=0$ ($k \ge 0$).
In Figs. \ref{fig_TR_T_local}(b) and (c), the resulting coefficients for $\epsilon/d=10^{-4}$ and  two different values
of $\alpha$ are shown. These figures have been calculated by
numerically solving the Schr\"odinger equation exactly and demonstrate that
$\alpha$ can indeed  be adjusted so
that $\fabsq{R^l} \approx 1$.
Fig. \ref{fig_TR_T_local}(c) demonstrates that the local PT-symmetric  potential works
as intended, i.e., as a transparent one-way reflector,  for a broad range of $k$ values.

{\it{Discussion.}}
This paper brings to the fore the essential role of eight generalized symmetries to determine the
transmission and reflection asymmetries by complex, and possibly nonlocal potentials.
These symmetries are classified with the aid of the relations between
unitary or antiunitary operators $1, \Pi, \Theta, \Theta\Pi$, which form Klein's 4-group,
and $H$ or its adjoint.
The symmetries set equalities among the scattering
amplitudes which, complemented by generalized unitarity relations, tell us which symmetries allow or disallow
a certain device with asymmetric scattering. Simplifying the analysis by imposing $0$ or $1$ scattering coefficients,
six possible device types exist.
We show how to design potentials realising these devices and provide examples on how to extend the domain of incident momenta
for which they work making use of Born's approximation.
The theory is worked out for particles and the Schr\"odinger equation but it is clearly of relevance for optical devices
due to the much exploited analogies and connections between Maxwell's equations and the Schr\"odinger equation,
which were used, e.g., to propose  the realization of PT-symmetric potentials in optics \cite{PT}.

Interesting questions left for future work are  the inclusion of other mechanisms for transmission and reflection asymmetries (for example
nonlinearities \cite{Xu,Konotop16}, and  time dependent potentials \cite{YuFan,Longhi2017}),
or a full discussion of the phases of the scattering amplitudes
in addition to the moduli emphasized here.
We shall also examine in a complementary paper the physical realization of complex nonlocal effective potentials.
In a quantum optics scenario, simple examples were provided in \cite{EPL} based on applying the partitioning technique \cite{F1,F2}
to the scattering of a particle with internal structure.

{\it{Acknowledgments}.}
We acknowledge financial support by the
Basque Government (Grant No.  IT986-16) and MINECO/FEDER,UE (Grant No. FIS2015-67161-P).
TD acknowledges support by the Irish Research Council (GOIPG/2015/3195).
\\
%

%
%

\pagebreak
\newpage
\clearpage

\begin{widetext}
\begin{center}
\textbf{ \large Supplemental Material: \\ Asymmetric scattering by non-hermitian potentials}
\end{center}
\end{widetext}

\setcounter{equation}{0} \setcounter{figure}{0} \setcounter{table}{0}
\setcounter{page}{6} \makeatletter \global\long\def\theequation{S\arabic{equation}}
\global\long\def\thefigure{S\arabic{figure}}
\global\long\def\thetable{S\Roman{table}}
\global\long\def\bibnumfmt#1{[S#1]}
\global\long\def\citenumfont#1{S#1}

\section{I. Scattering amplitudes\label{sa}}
We provide a lightning review of scattering amplitudes in 1D. For a more complete account, see [1].
We assume $p>0$. The amplitudes for scattering by $H=H_0+V$, may be calculated by
\beqa
R^l&=&-\frac{2\pi i m}{p}\la -p|T_{op}(+)|p\ra,
\\
T^l&=&1-\frac{2\pi i m}{p} \la p|T_{op}(+)|p\ra,
\label{tl}
\\
R^r&=&-\frac{2\pi i m}{p}\la p|T_{op}(+)|-p\ra,
\\
T^r&=&1-\frac{2\pi i m}{p}\la -p|T_{op}(+)|-p\ra,
\label{tr}
\eeqa
where the $l/r$ superscript indicates left or right incidence, and
\beq
T_{op}(+)|\pm p\ra=\left[V+V\frac{1}{E_p+i0-H}V\right]|\pm p\ra,
\eeq
where $E_p=p^2/(2m)$.
To find Born-approximation expressions of the scattering coefficients (square moduli of the amplitudes), we take $T_{op}\approx V$ in the expressions of $R^l$, and $R^r$.
For $T^l$ and $T^r$ we also include the second order in $V$, which contributes to the square in
second order due to the $1$ in Eqs. (\ref{tl}) and (\ref{tr}).

\begin{figure}
\includegraphics[width = 0.9\columnwidth]{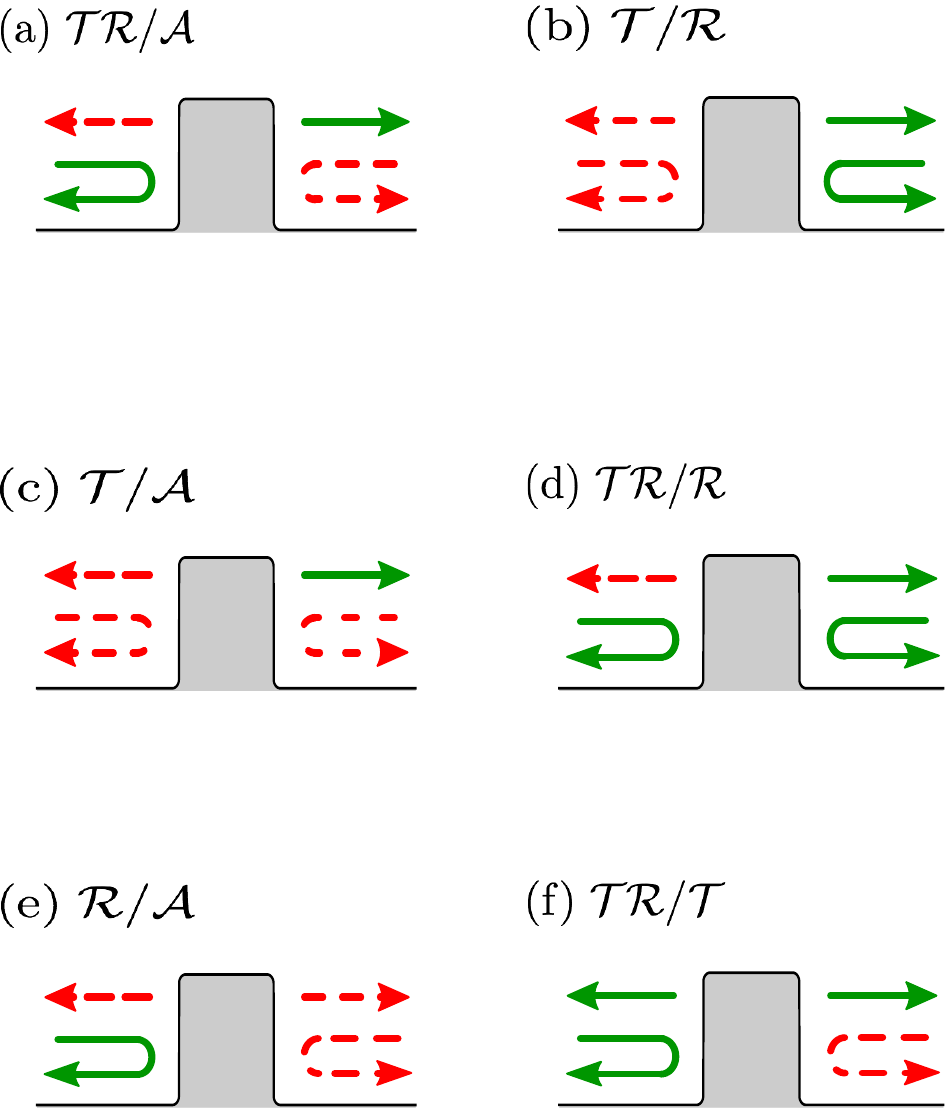}
\caption{(Color online) Schemes for extreme asymmetric transmission.  The dashed and continuous lines represent respectively zero or one
for the moduli of the scattering amplitudes; the bended lines are for reflection amplitudes, and the straight lines for transmission:
(a) One-way mirror ($\cal{TR/A}$); (b) One-way barrier ($\cal{T/R}$); (c) One-Way T-filter ($\cal{T/A}$);
(d) Mirror \& 1-way transmitter ($\cal{TR/R}$); (e) One-way R-filter ($\cal{R/A}$); (f) Transparent, one-way
reflector ($\cal{TR/T}$)
\label{cases}}
\end{figure}


\begin{figure}[h]
\begin{center}
(a)\includegraphics[width = 0.6\columnwidth]{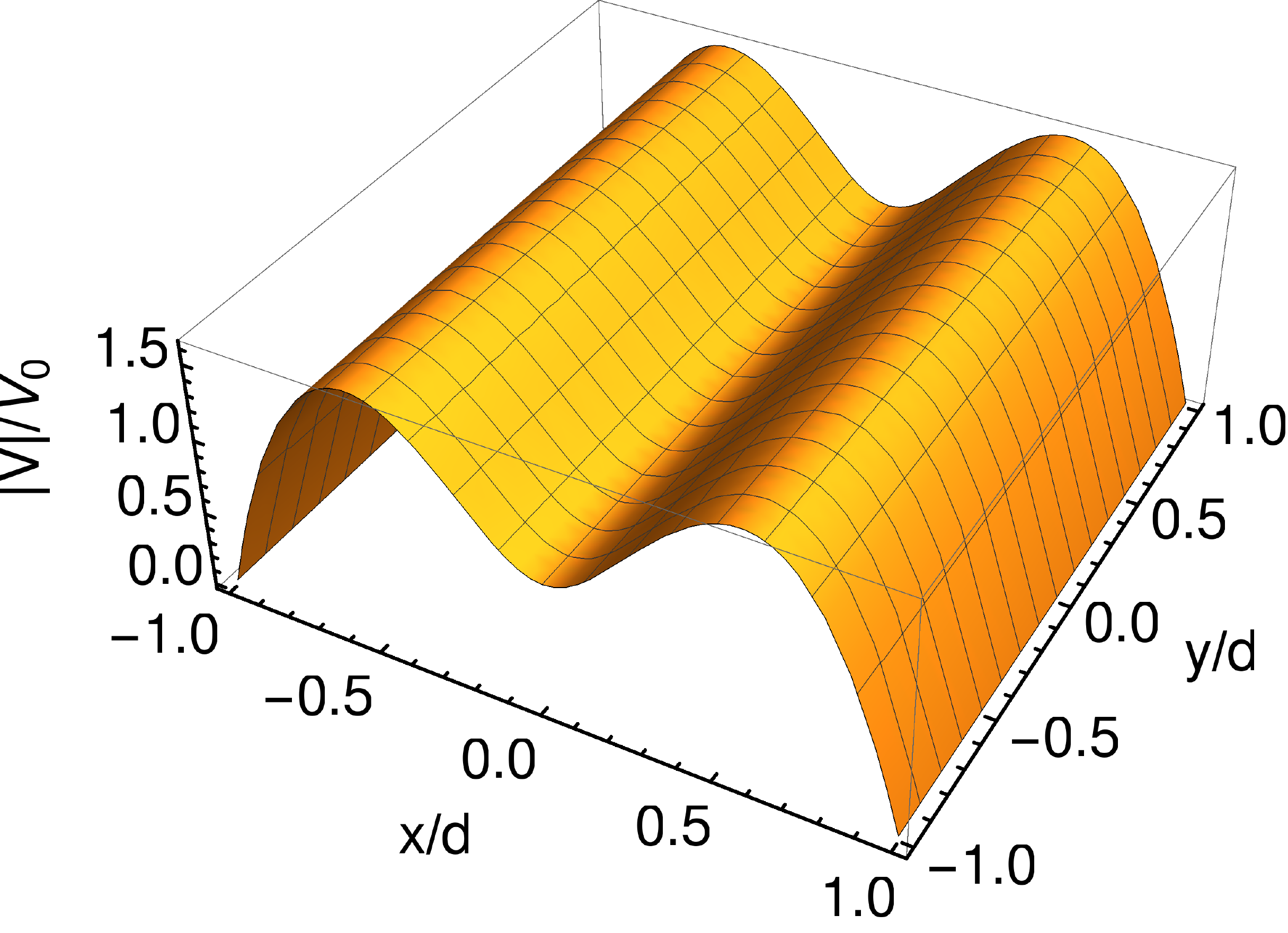}\\
(b)\includegraphics[width = 0.6\columnwidth]{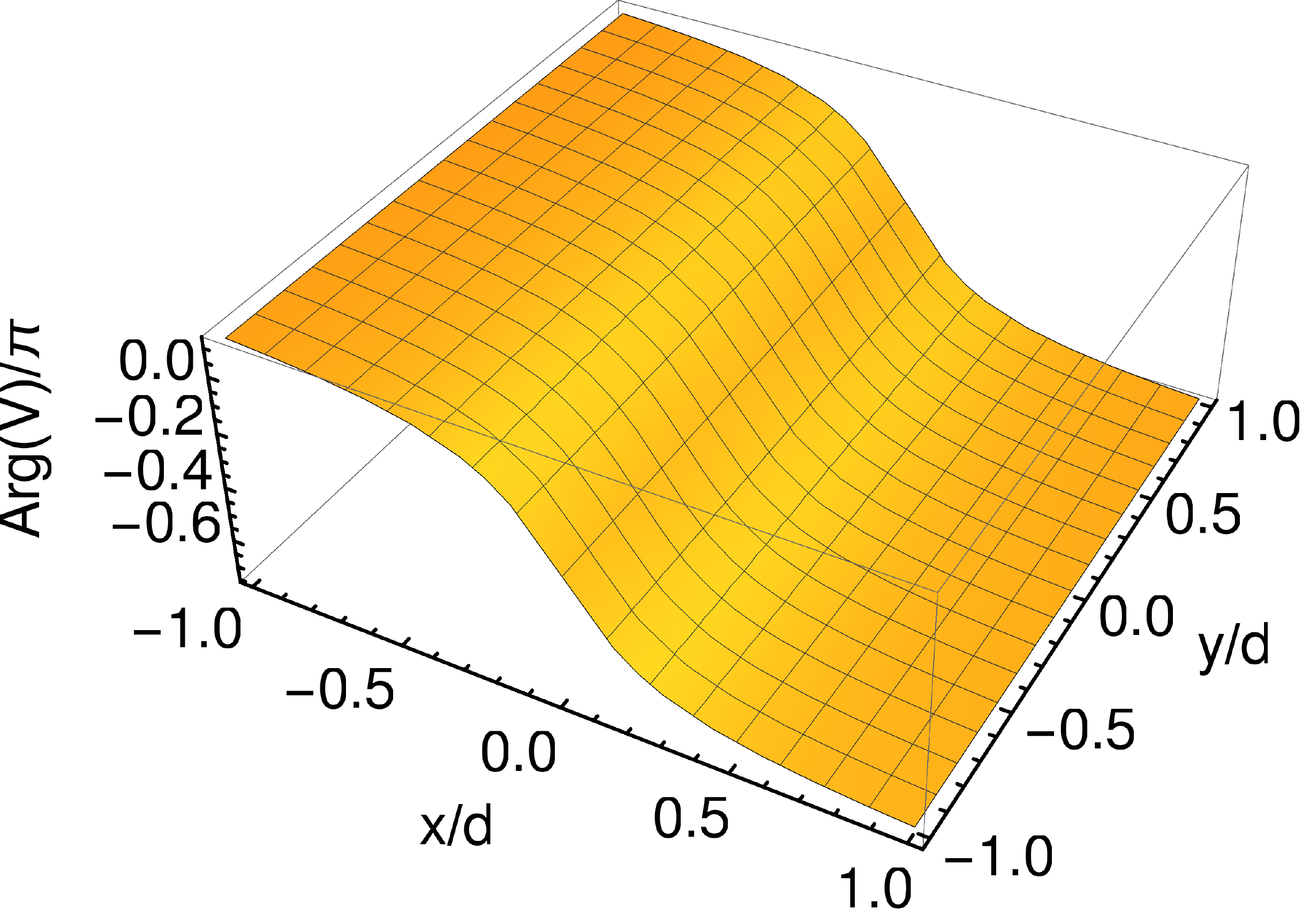}\\
(c)\includegraphics[width = 0.6\columnwidth]{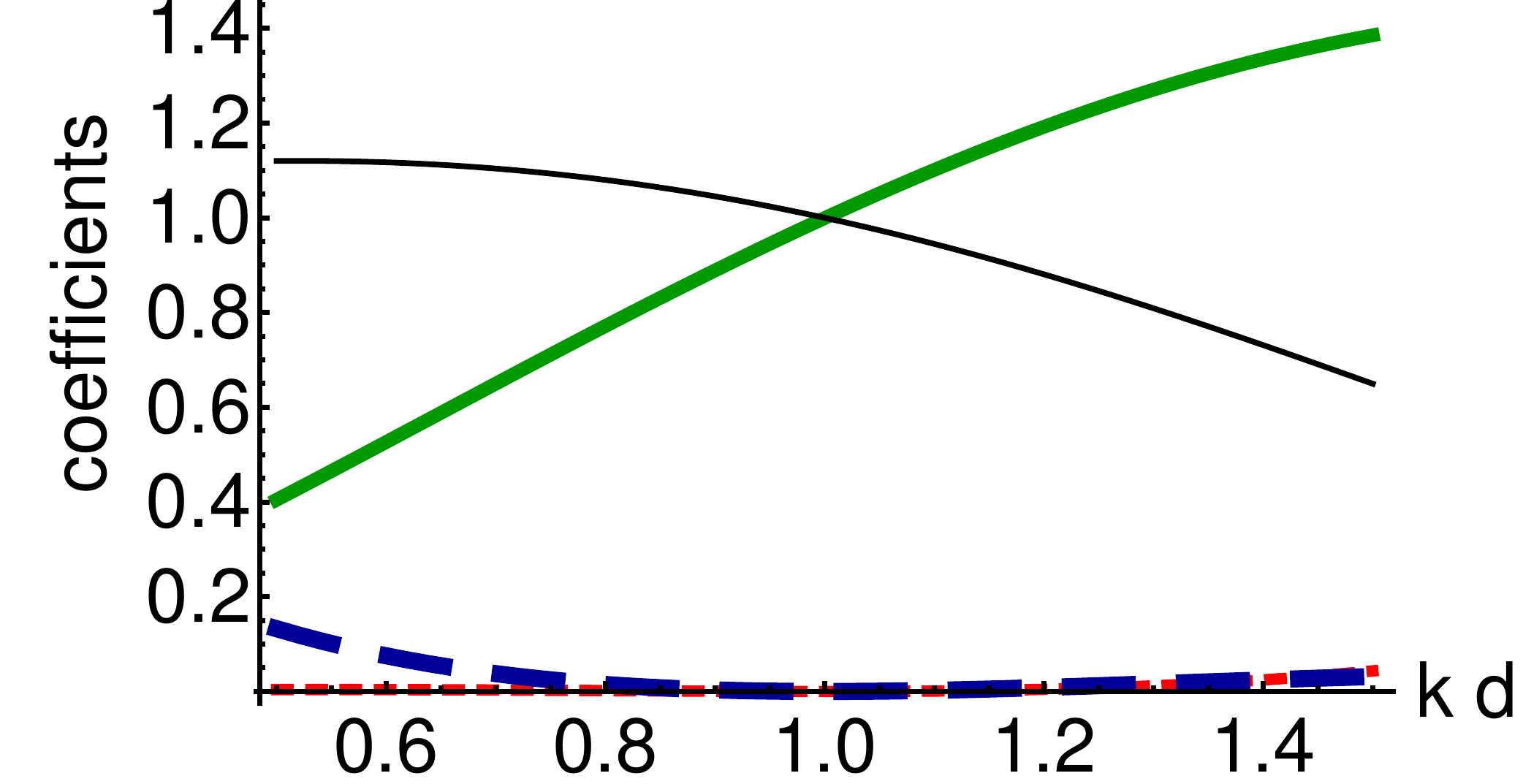}
\end{center}
\caption{(Color online) One-way mirror ($\cal{TR/A}$, $R^l=-1, R^r=0$): potential $V(x,y)=|V(x,y)|e^{i\Phi}$ for $k_0 = 1/d$;
(a) absolute value $\fabs{V(x,y)}$, (b) argument $\Phi$,
(c) transmission and reflection coefficients,
left incidence: $\fabsq{R^l}$ (black, solid line), $\fabsq{T^l}$ (green, solid line);
right incidence: $\fabsq{R^r}$ (blue, tick, dashed line), $\fabsq{T^r}$ (red, dotted line).\label{fig_device_TR_A}}
\end{figure}


\begin{figure}[h]
\begin{center}
(a)\includegraphics[width = 0.6\columnwidth]{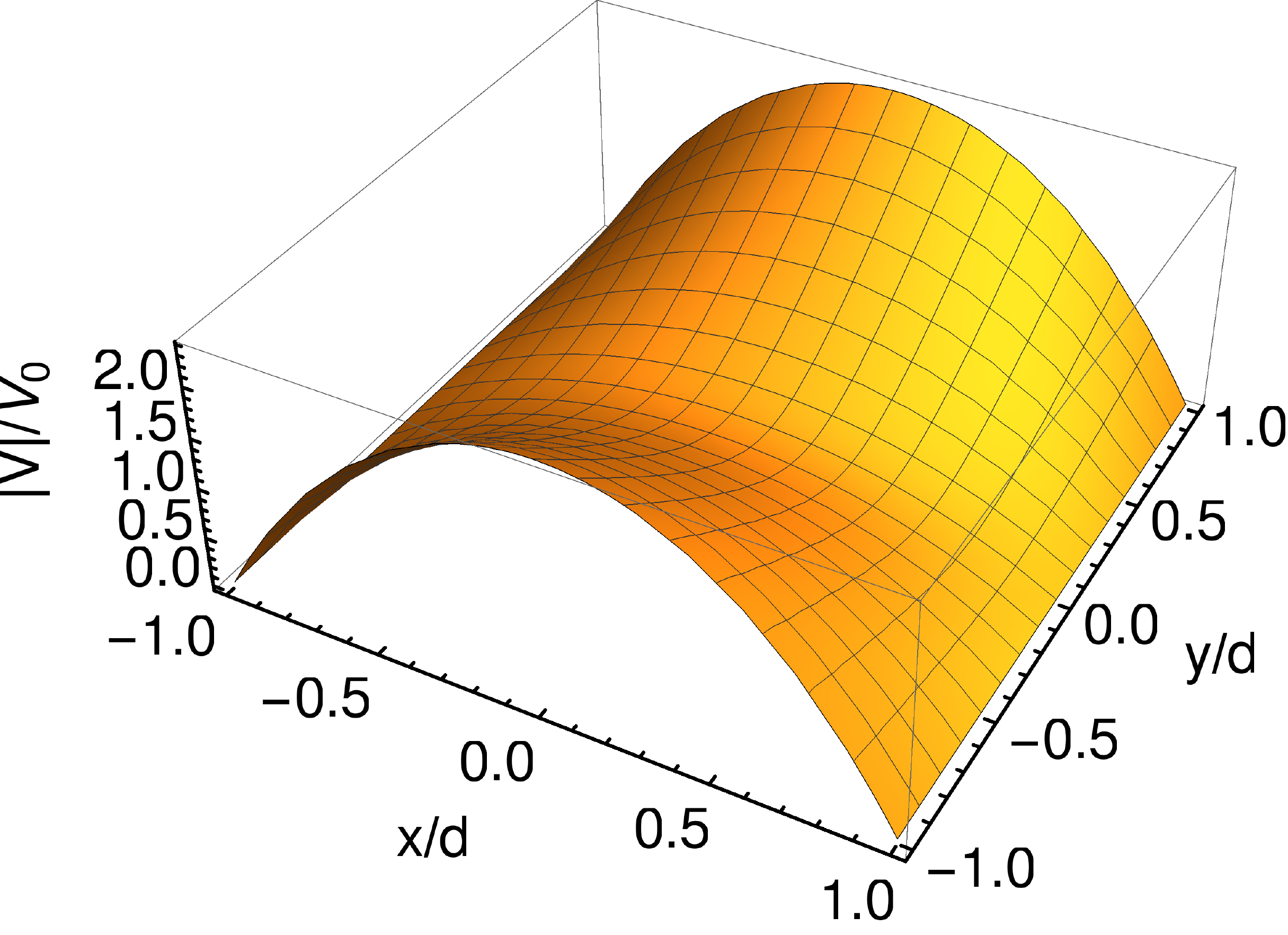}\\
(b)\includegraphics[width = 0.6\columnwidth]{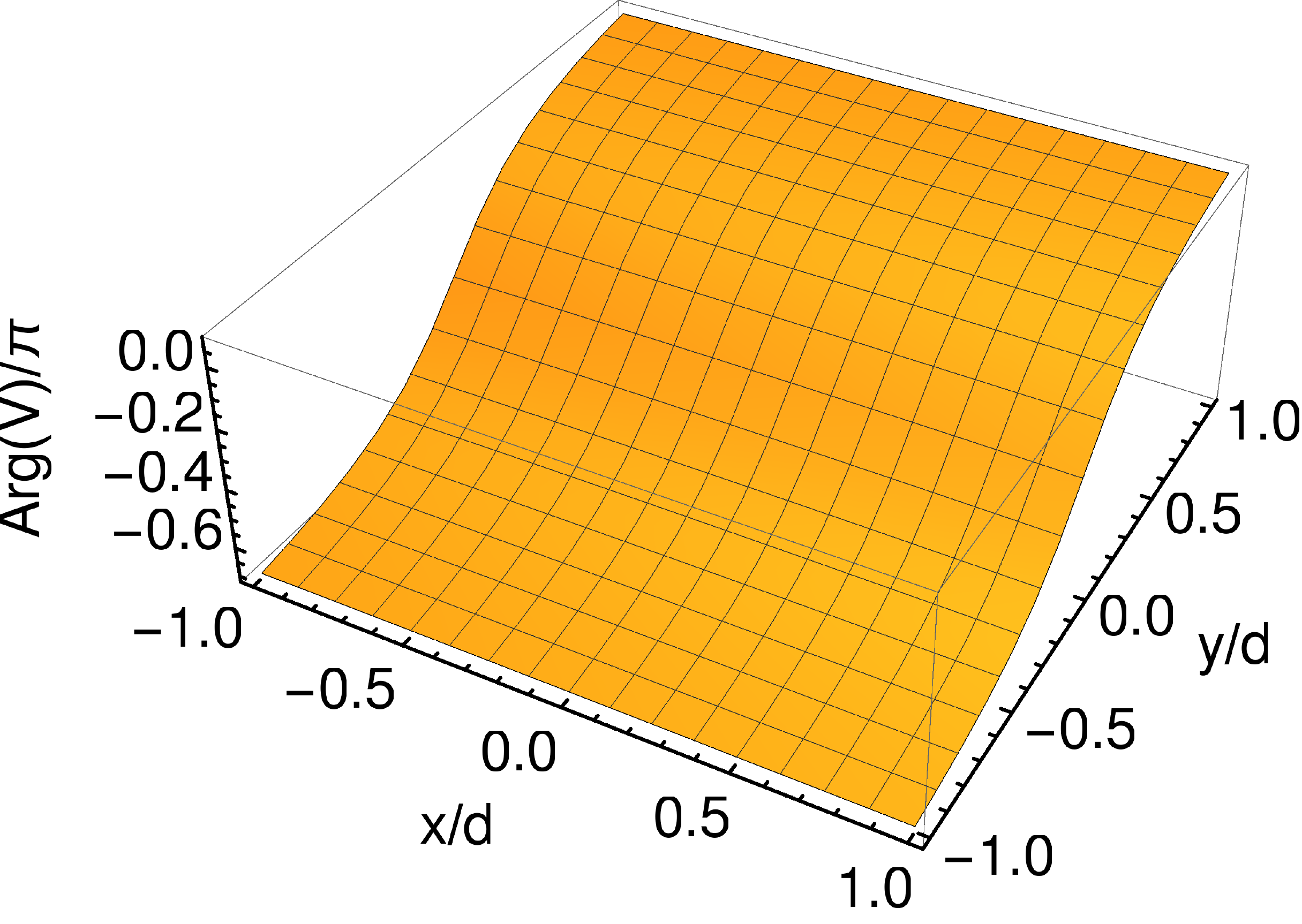}\\
(c)\includegraphics[width = 0.6\columnwidth]{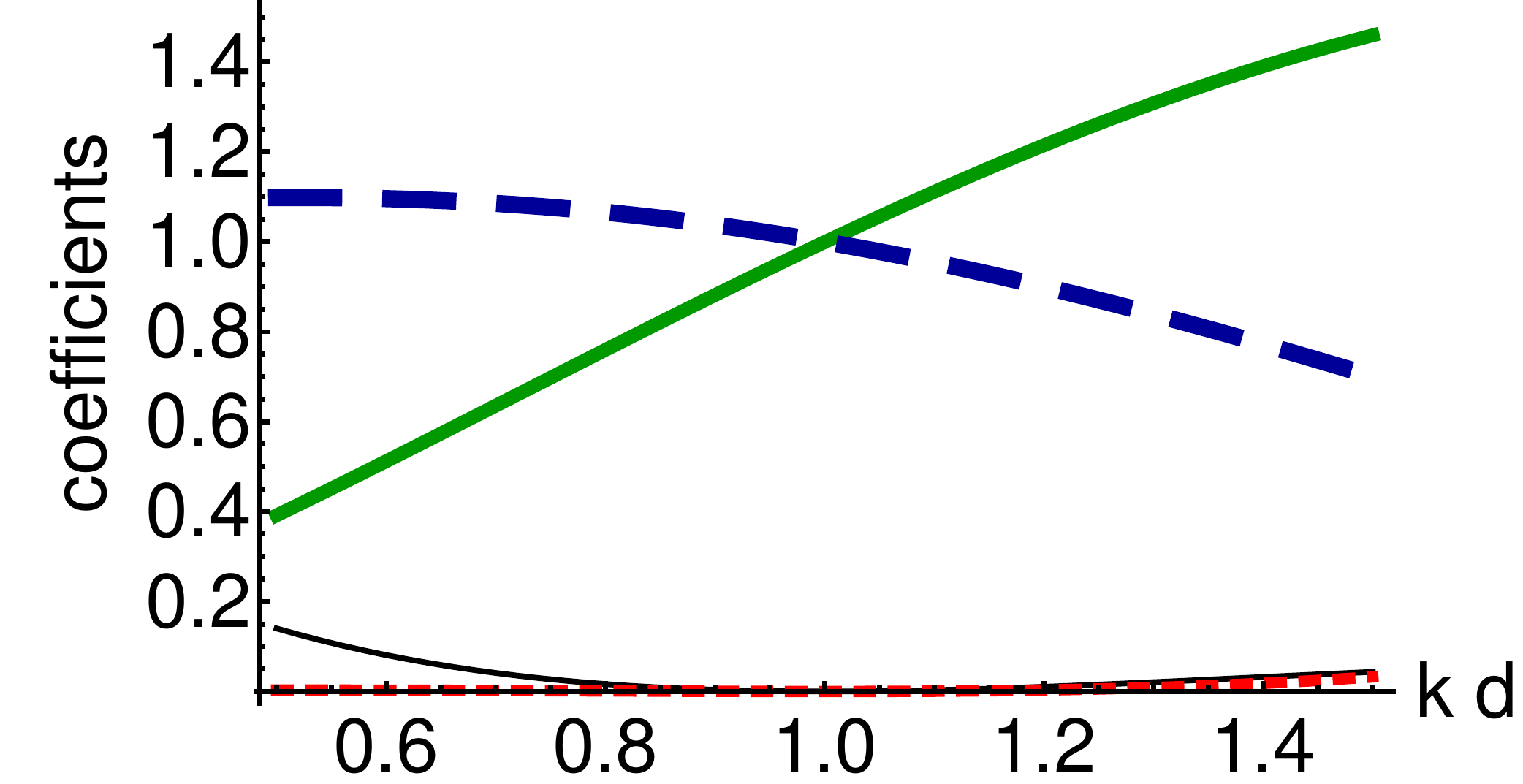}
\end{center}
\caption{(Color online) One-way barrier ($\cal{T/R}$, $R^l=0, R^r=-1$): potential $V(x,y)=|V(x,y)|e^{i\Phi}$ for $k_0 = 1/d$;
(a) absolute value $\fabs{V(x,y)}$, (b) argument $\Phi$,
(c) transmission and reflection coefficients,
left incidence: $\fabsq{R^l}$ (black, solid line), $\fabsq{T^l}$ (green, solid line);
right incidence: $\fabsq{R^r}$ (blue, tick, dashed line), $\fabsq{T^r}$ (red, dotted line).\label{fig_device_T_R}}
\end{figure}

The on-shell $\sf{S}$ matrix, see [1],  is formed
as
\beq
\sf{S}=\left(
\begin{array}{cc}
\la \bf{p}|\sf{S}|\bf{p}\ra&\la \bf{p}|\sf{S}|-\bf{p}\ra
\\
\la -�\bf{p}|\sf{S}|\bf{p}\ra&\la -\bf{p}|\sf{S}|-\bf{p}\ra
\end{array}
\right)
=
\left(\begin{array}{cc}
T^l&R^r
\\
R^l&T^r
\end{array}
\right).
\eeq
This on-shell matrix relates to the standard $S$-matrix elements
in momentum representation,
\beq
\la p|S|p'\ra=\delta(p-p')-2i\pi\delta(E_p-E_p')\la p|T_{op}(+)|p'\ra,
\eeq
by factoring out a delta function,
%
$
\la p|S|p'\ra=\frac{|p|}{m}\delta(E_p-E_p')\la \bf{p}|\sf{S}|\bf{p'}\ra.
$
%
All the above formulae may be reproduced when the  particle is scattered instead by  $H^\dagger=H_0+V^\dagger$,
giving scattering amplitudes with a hat, $\widehat{T}^r, \widehat{T}^l, \widehat{R}^r, \widehat{R}^l$, and $\widehat{S}$.
Hatted and unhatted amplitudes are not independent, they are linked by the generalized unitary relation $\widehat{S}^\dagger S=S\widehat{S}^\dagger=1$, whose
on-shell matrix elements lead to the four relations
\begin{eqnarray}
\widehat T^l T^{l*} + \widehat R^l R^{l*} = 1,
\nonumber\\
\widehat T^r T^{r*} + \widehat R^r R^{r*} = 1,
\nonumber\\
\hat T^{l*} R^r + T^r \widehat R^{l*} = 0,
\nonumber\\
T^l \widehat R^{r*} + \widehat T^{r*} R^l = 0.
\label{gur}
\end{eqnarray}
They can be rearranged to express the transmission amplitudes of $H^\dagger$ in terms of   those of $H$,
\begin{eqnarray}
\widehat T^{l*} = \frac{T^r}{T^l T^r - R^l R^r},\;
\widehat R^{l*} = - \frac{R^r}{T^l T^r - R^l R^r},\nonumber\\
\widehat T^{r*} = \frac{T^l}{T^l T^r - R^l R^r},\;
\widehat R^{r*} = - \frac{R^l}{T^l T^r - R^l R^r}.
\label{hadjamp}
\end{eqnarray}
\section{II. Examples of potentials for devices with asymmetric-scattering coefficients}
\subsection{IIa. Nonlocal potentials for devices with transmission asymmetry}
The asymmetric-transmission devices
($|T^l| = 1, |T^r|=0$, $|R^{r,l}|=0,1$) can be seen in Fig. \ref{cases} (a,b,c,d).
For constructing examples of potentials for such devices, we  fix the phases of the transmission
amplitudes as $T^l = 1, T^r=0$, and the reflection amplitudes
will be specified in each case.  We assume the form $V (x, y) = \sum_{i=0}^5 \sum_{j=0}^1 v_{ij} x^i y^j$
for the potential,  plug this ansatz in the Schr\"odinger equation (3),
and equate equal powers of $x$.
Moreover we demand that $V(-d,y)=0 = V(d,y)$ for all $y$ such that the total potential
(including the vanishing potential for $x,y<-d$ and $x,y>d$) is continuous.

We consider first an ideal  one-way mirror ($\cal{TR/A}$) with amplitudes
$R^l = -1, R^r = 0$.
Waves sent from  the left are fully reflected, but there is also perfect transmission, whereas
waves sent from the right are absorbed.
The potential that achieves this for $k = k_0= 1/d$
is shown in Figs. \ref{fig_device_TR_A}(a),(b) where $V_0 = \hbar^2/(2m d^3)$.
Similarly, the potential of a one-way barrier ($\cal{T/R}$) is shown in Figs. \ref{fig_device_T_R} (a),(b)
with $R^l = 0, R^r=-1$.
Note that the potential matrices or potential kernel functions
$V(x,y)$  do not have units of energy but units of a force.
In agreement with Table III,
these  potentials do not satisfy any of the nontrivial symmetries II, III, ...,VIII.
The transmission and reflection coefficients around $k_0$ are also depicted in Figs. \ref{fig_device_TR_A} (c) and \ref{fig_device_T_R} (c), which show that the desired values are
achieved exactly at $k_0$ but also approximately in some neighborhood of $k_0$. This  holds true  for all
potentials in this Supplemental Material.
%
%
\subsection{IIb. Nonlocal potentials fulfilling symmetry VIII for devices with transmission asymmetry}
One-way T-filters ($\cal{T/A}$) and the mirror\&1-way transmitters ($\cal{TR/R}$) can be also
constructed using the method described in the previous subsection.
Nevertheless, unlike the two devices in the previous subsection, these devices can fulfill symmetry VIII.
We  assume now the form $V (x, y) = \sum_{i=0}^5 \sum_{j=0}^5 v_{ij} x^i y^j$
with $v_{ij} = (-1)^{i+j} v_{ji}$.
To  simplify the potential, we also demand $v_{4,4}=v_{4,5} = v_{5,4}=v_{5,5}=0$.
Moreover we demand that $V(-d,y)=0 = V(d,y)$ for all $y$ such that the total potential
(including the vanishing potential for $x,y<-d$ and $x,y>d$) is continuous.
It is also required that  $R^l=R^r=R$, consistent with Table I.

\begin{figure}[t]
\begin{center}
(a)\includegraphics[width = 0.6\columnwidth]{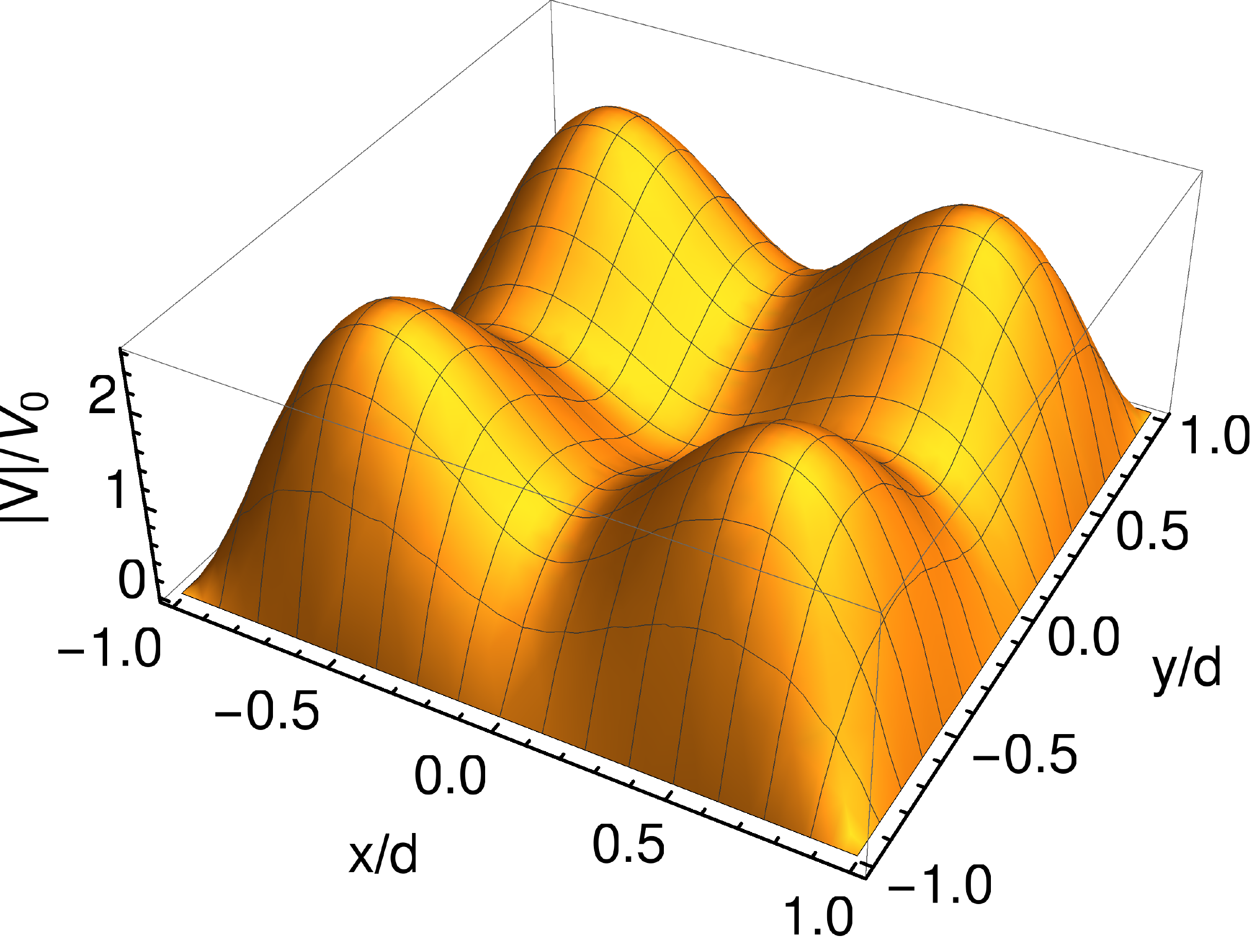}\\
(b)\includegraphics[width = 0.6\columnwidth]{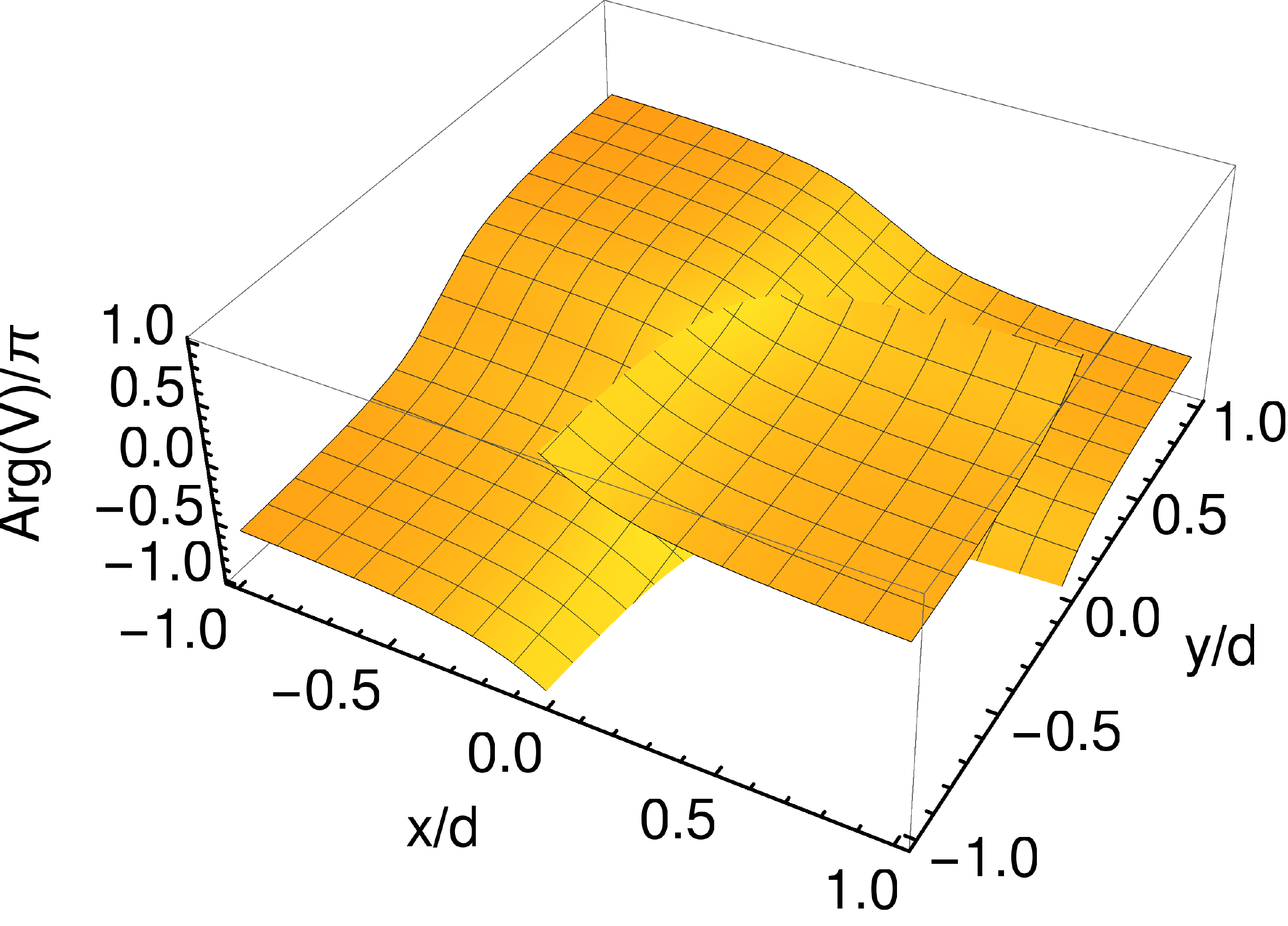}\\
(c)\includegraphics[width = 0.6\columnwidth]{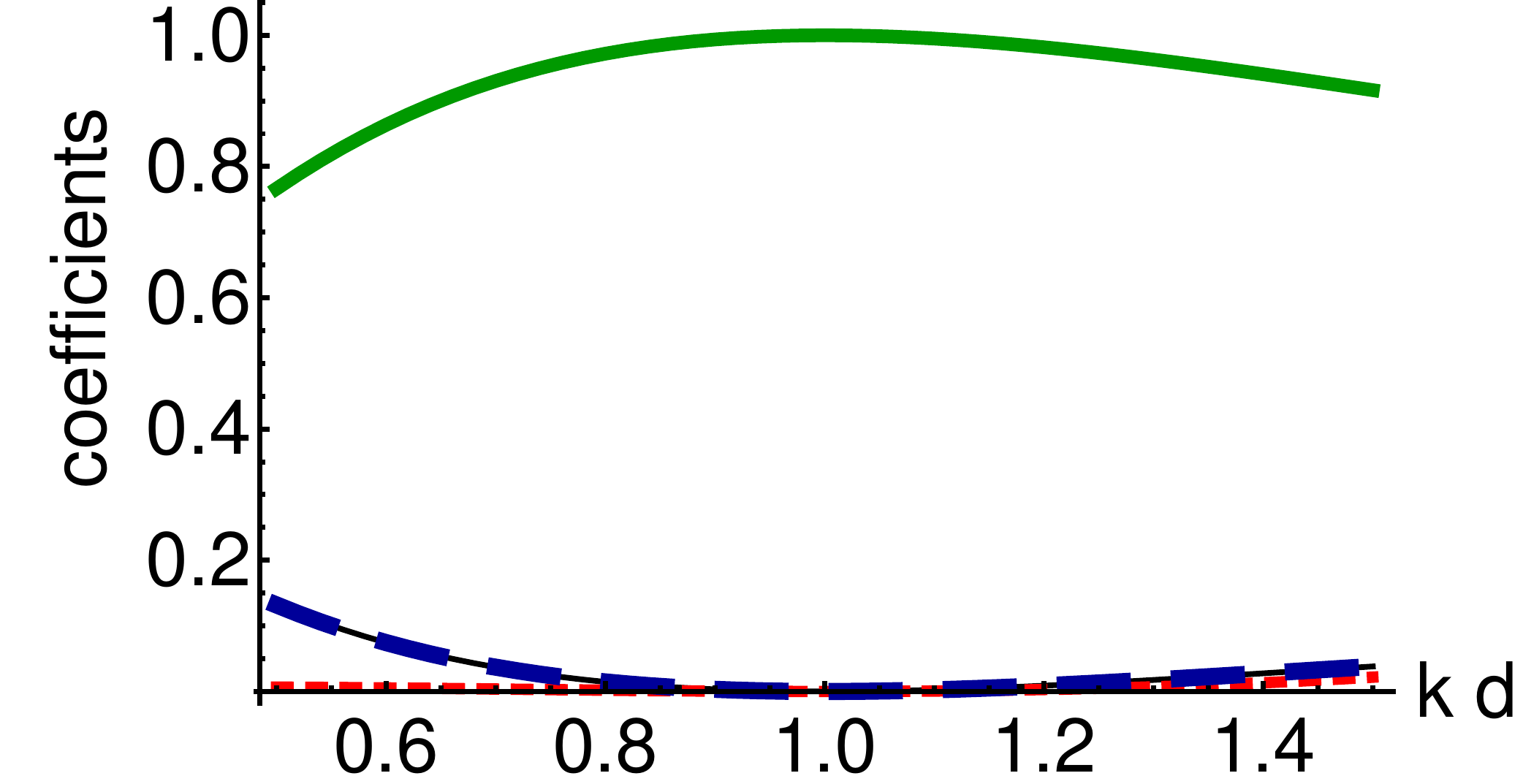}
\end{center}
\caption{(Color online) One-way T-filter ($\cal{T/A}$, $R^l=0, R^r=0$): potential $V(x,y)=|V(x,y)|e^{i\Phi}$ for $k_0 = 1/d$;
(a) absolute value $\fabs{V(x,y)}$, (b) argument $\Phi$,
(c) transmission and reflection coefficients,
left incidence: $\fabsq{R^l}$ (black, solid line), $\fabsq{T^l}$ (green, solid line);
right incidence: $\fabsq{R^r}$ (blue, tick, dashed line), $\fabsq{T^r}$ (red, dotted line).\label{fig_device_T_A}}
\end{figure}

\begin{figure}[ht]
\begin{center}
(a)\includegraphics[width = 0.6\columnwidth]{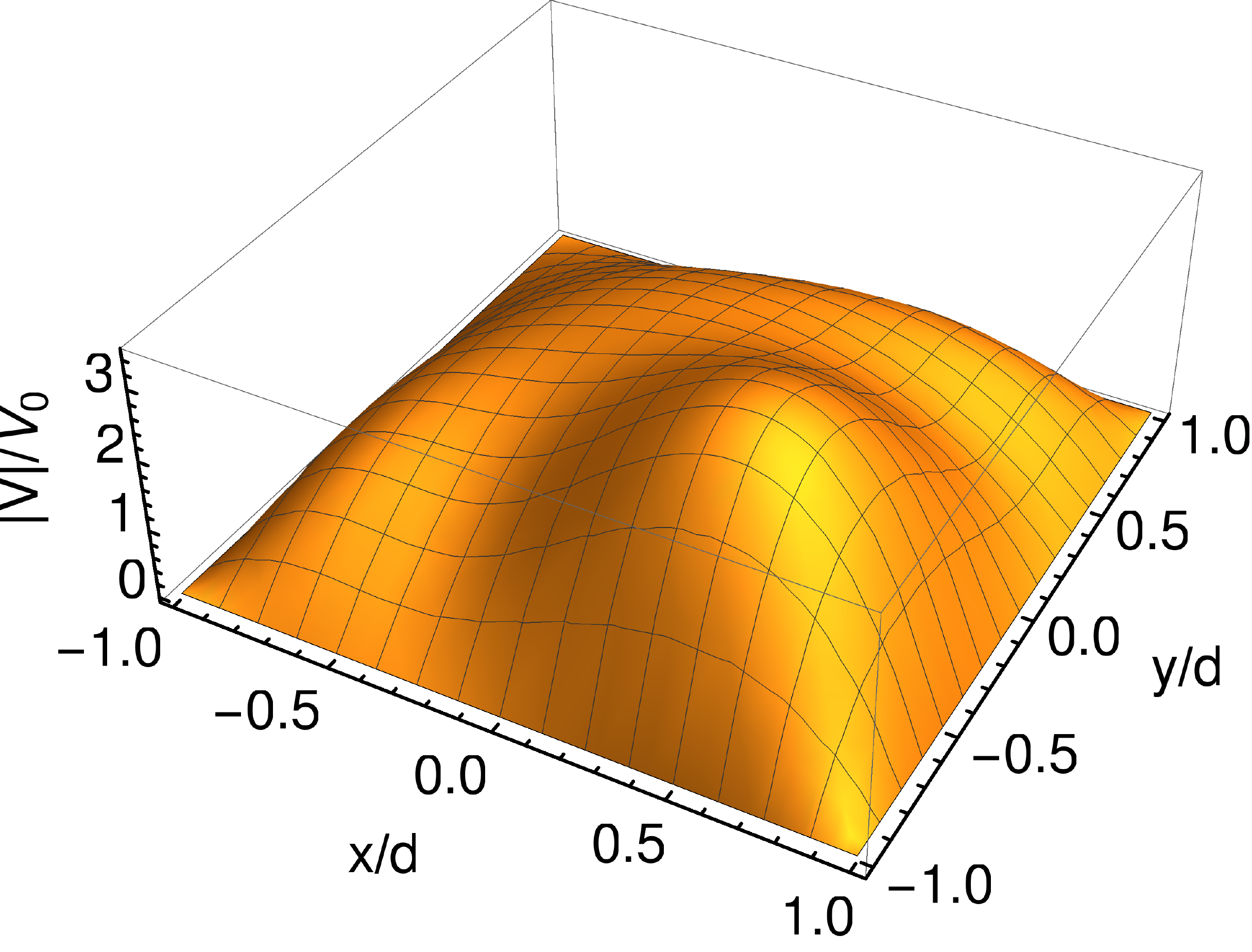}\\
(b)\includegraphics[width = 0.6\columnwidth]{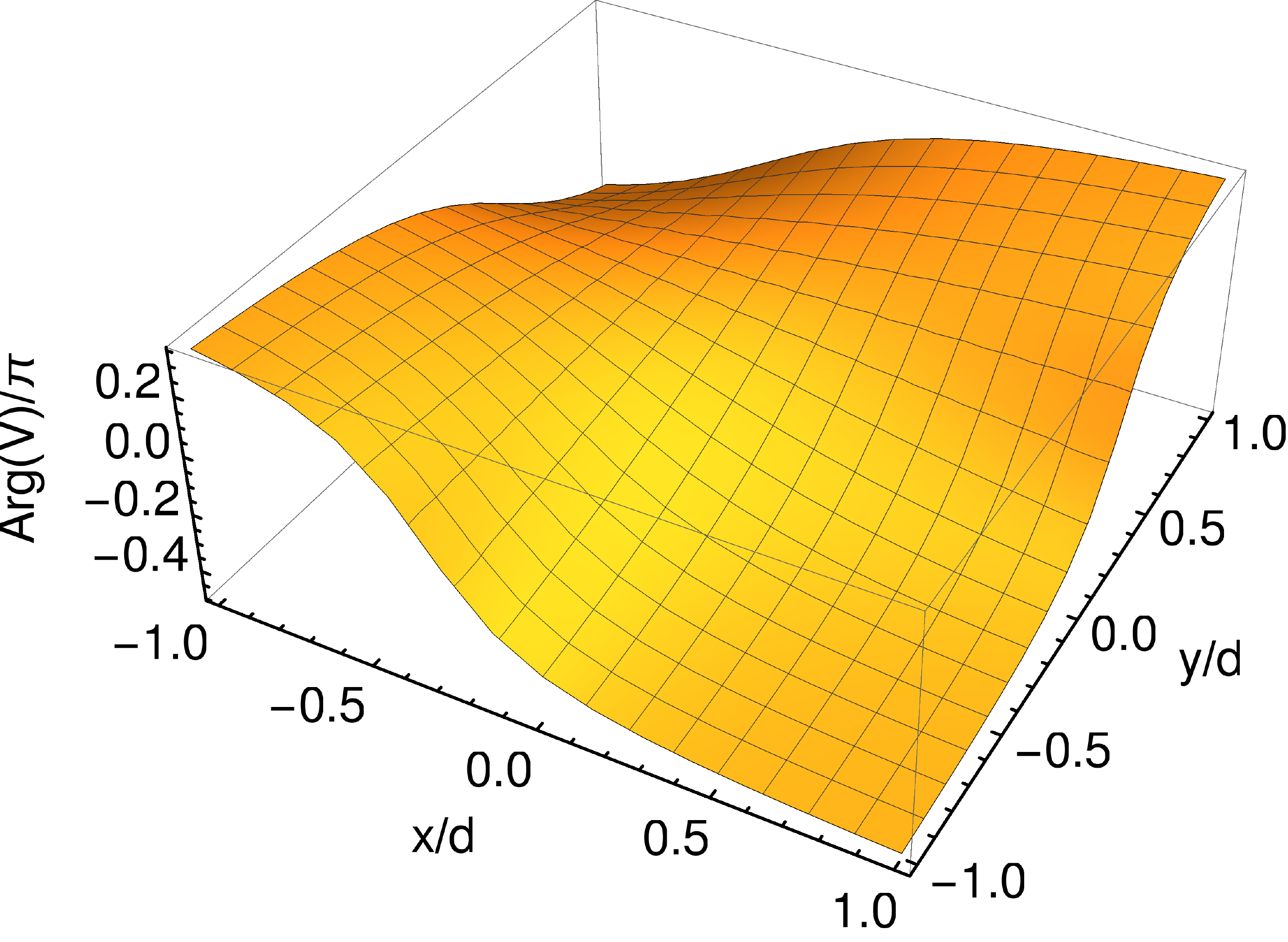}\\
(c)\includegraphics[width = 0.6\columnwidth]{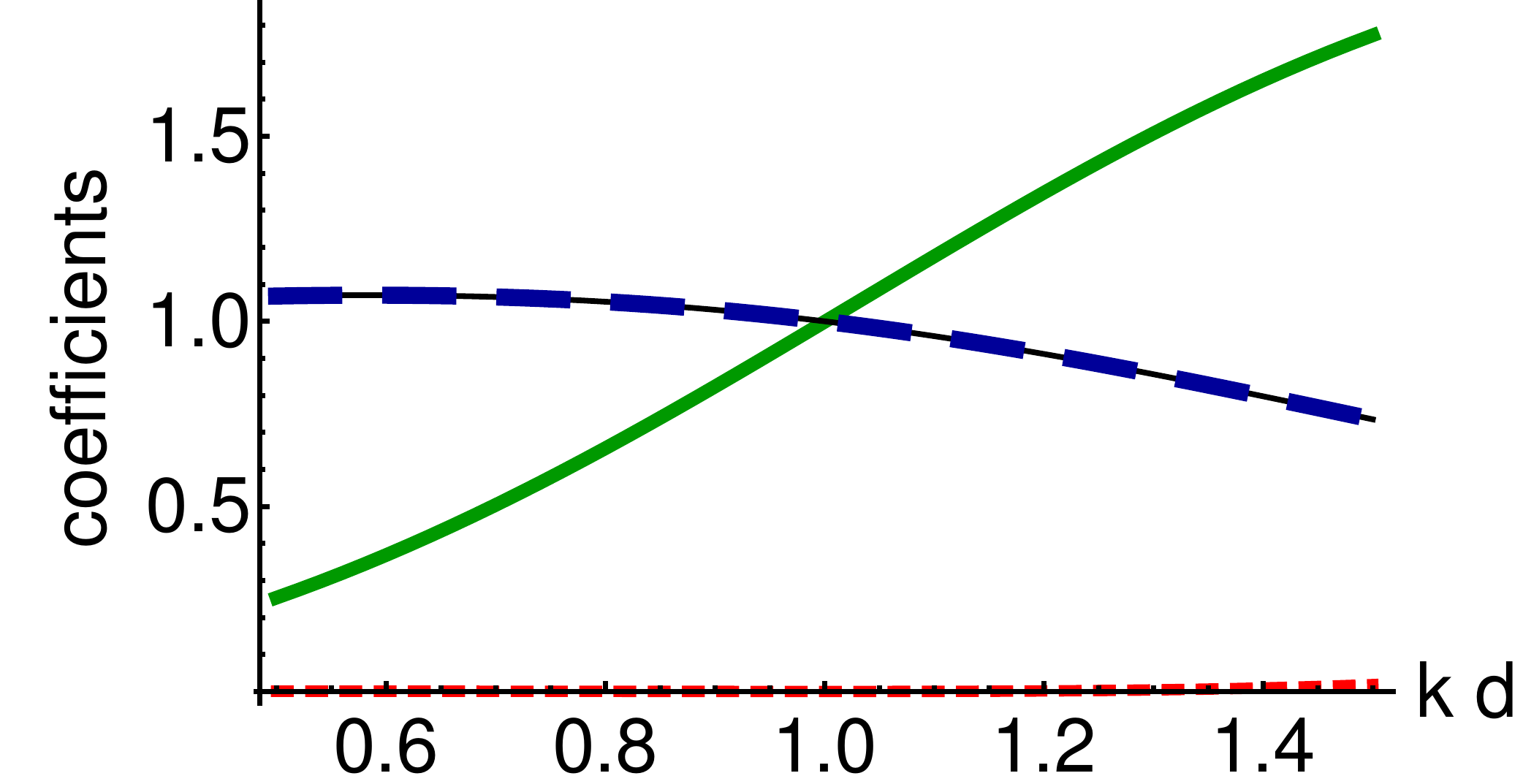}
\end{center}
\caption{(Color online) Mirror\&1-way transmitter ($\cal{TR/R}$, $R^l=-1, R^r=-1$): potential $V(x,y)=|V(x,y)|e^{i\Phi}$ for $k_0 = 1/d$;
(a) absolute value $\fabs{V(x,y)}$, (b) argument $\Phi$,
(c) transmission and reflection coefficients,
left incidence: $\fabsq{R^l}$ (black, solid line), $\fabsq{T^l}$ (green, solid line);
right incidence: $\fabsq{R^r}$ (blue, tick, dashed line), $\fabsq{T^r}$ (red, dotted line).\label{fig_device_TR_R}
}
\end{figure}

In Fig. \ref{fig_device_T_A}, the potential for the one-way T-filter ($\cal{T/A}$), with $R=0, T^l=1$, is shown,
and the potential for the mirror\&1-way transmitter ($\cal{TR/R}$), calculated for  $R=-1, T^l=1$,
is shown in Fig. \ref{fig_device_TR_R} where we have chosen $k_0= 1/d$.
The transmission and reflection coefficients around $k_0$ are also depicted.

For the first three devices ($\cal{TR/A}, \cal{T/R}$ and $\cal{T/A}$),
it follows from the  generalized  unitarity relations \eqref{hadjamp} that
one or more of the transmission and reflection amplitudes of the corresponding
adjoint Hamiltonian will diverge at $k=k_0=1/d$ (if the numerator
on the right-hand side of these relations stays finite while the corresponding
denominator $T^l T^r - R^l R^r= - R^l R^r \to 0$).
In the mirror\&1-way transmitter, it follows from \eqref{hadjamp} that
$\widehat T^l = 0, \widehat R^l = -1, \widehat T^r = -1, \widehat R^r = -1$, and therefore
the adjoint Hamiltonian provides a mirror\&1-way transmitter device with $l \leftrightarrow r$.


\begin{figure}[t]
\begin{center}
(a)\includegraphics[width = 0.6\columnwidth]{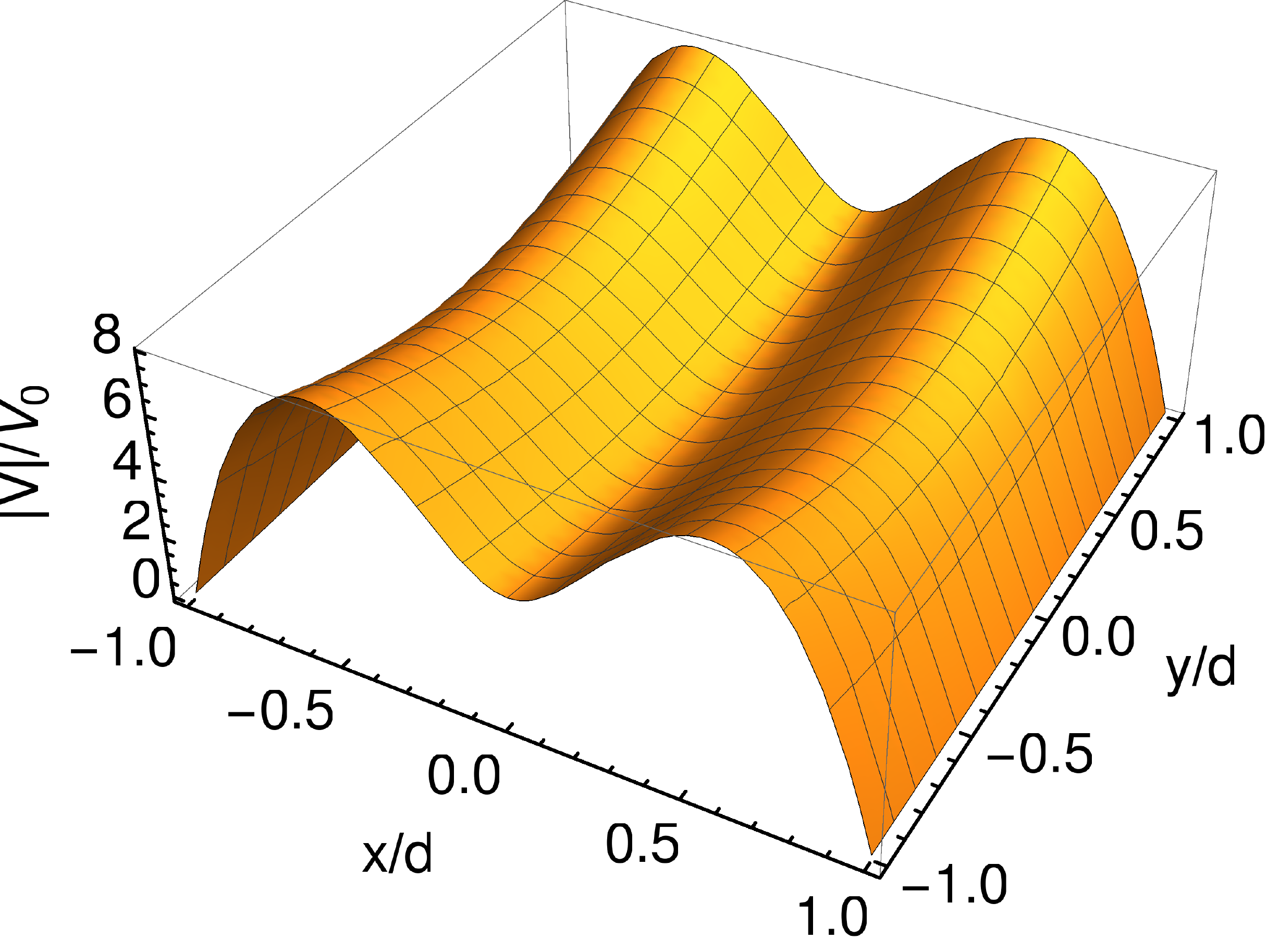}\\
(b)\includegraphics[width = 0.6\columnwidth]{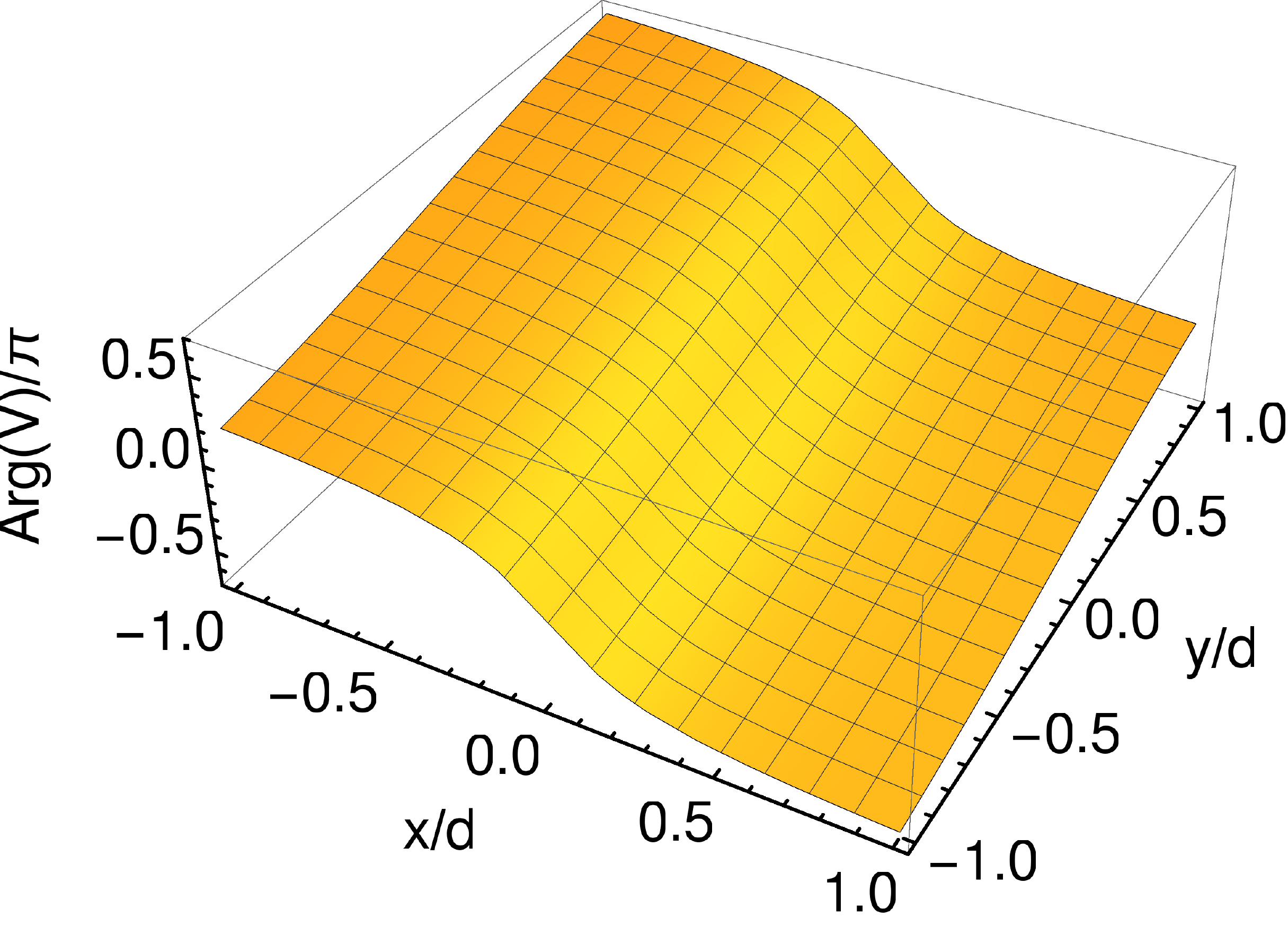}\\
(c)\includegraphics[width = 0.6\columnwidth]{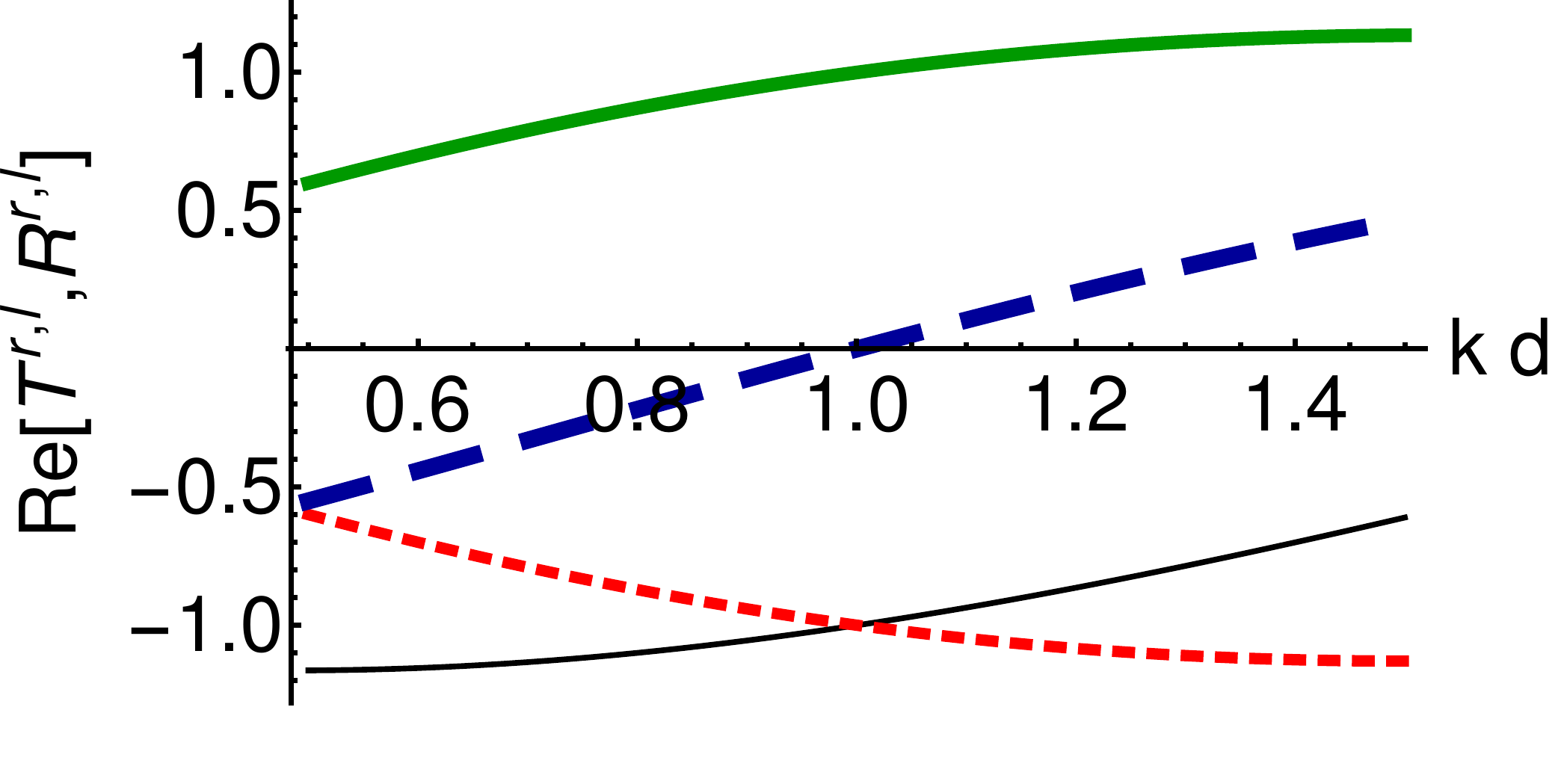}\\
(d)\includegraphics[width = 0.6\columnwidth]{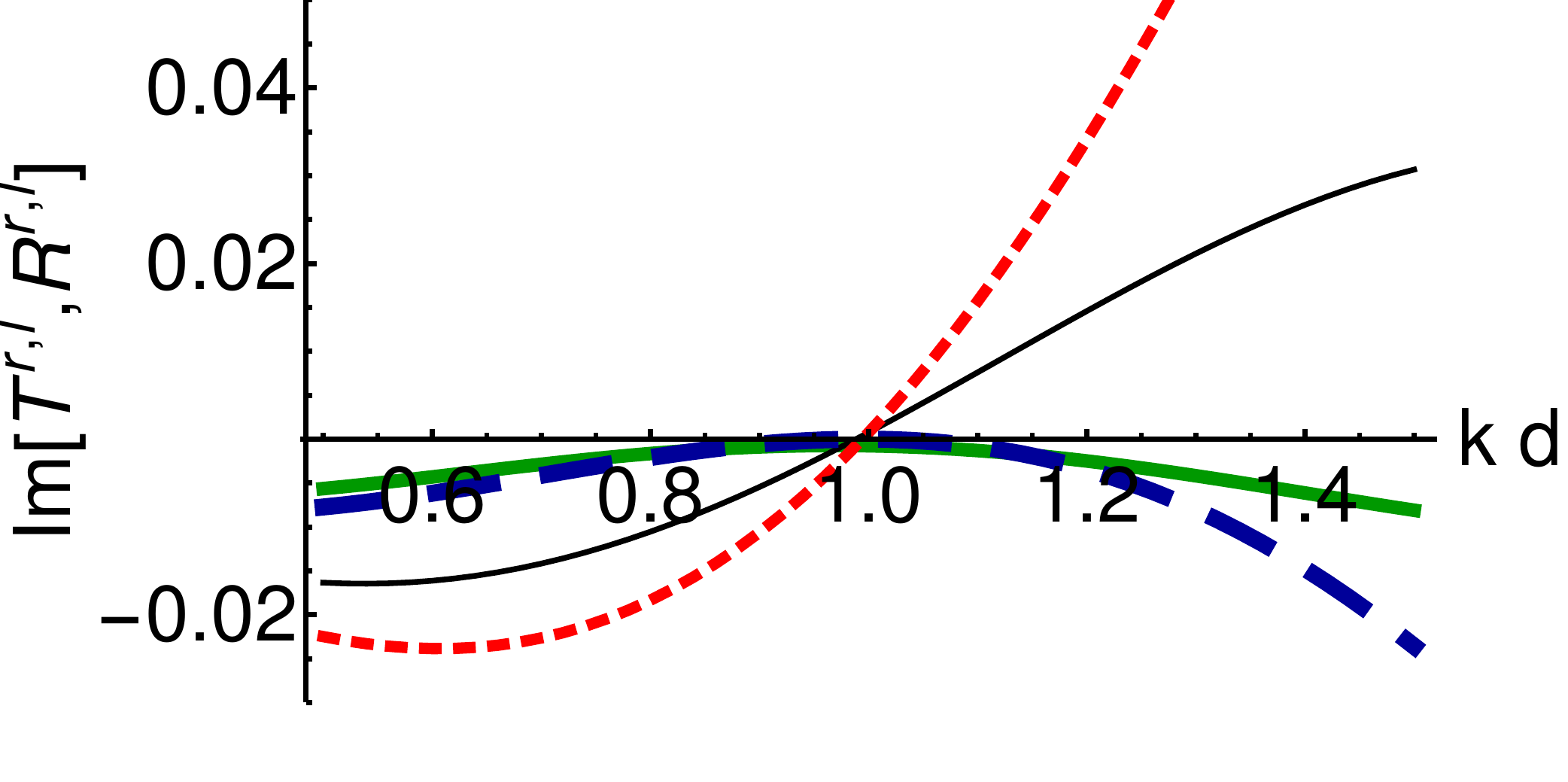}
\end{center}
\caption{(Color online) Nonlocal PT-symmetric potential leading to $T^l =1, R^l = -1, T^r=-1, R^r = 0$:
potential $V(x,y)=|V(x,y)|e^{i\Phi}$ for $k_0 d=1$;
(a) absolute value $\fabs{V(x,y)}$, (b) argument $\Phi$,
(c) real and (d) imaginary part of the transmission and reflection amplitudes,
left incidence: ${R^l}$ (black, solid line), ${T^l}$ (green, solid line);
right incidence: ${R^r}$ (blue, tick, dashed line), ${T^r}$ (red, dotted line).
\label{fig_nonlocal_pt}
}
\end{figure}
%
%
\subsection{IIc. Devices with asymmetric reflection}
In the previous subsections we have already considered two device types with asymmetric reflection coefficients, namely,
the one-way mirror ($\cal{TR/A}$), and the one way-barrier ($\cal{T/R}$).
These are the only two device types which are
simultaneously asymmetrical for transmission and reflection.
Two more types are possible which have only reflection asymmetry, namely,  the one-way $R$-filter ($\cal{R/A}$), and the transparent one-way reflector
($\cal{TR/T}$). Both are compatible with symmetry type VI, in particular with local potentials.

\begin{table}
\centering
\scalebox{1.0}{
\begin{tabular}{|c|c|}
\hline
Symmetry& Allowed devices
\\
\hline
I&All types
\\
II&None
\\
III&None
\\
IV&$\cal{TR/R,TR/T}$
\\
V&$\cal{TR/R}$
\\
VI & $\cal{R/A, TR/T}$
\\
VII&$\cal{TR/T}$
\\
VIII&$\cal{T/A,TR/R}$
\\
\hline
\end{tabular}}
\caption{Device types allowed for a given symmetry.
\label{table3}}
\end{table}

A one-way $R$-filter $\cal{R/A}$ acts as a perfect absorber from one side and as a perfect reflector from the other side. It may thus be  constructed
by adding an infinite barrier with its edge touching the end of known-perfect absorbers for one-sided incidence [1,11-13].
Local, perfect absorbers can be worked out for one
or more incident momenta, or for a momentum window.
According to Table III,
a $\cal{R/A}$ device cannot have PT-symmetry. Indeed experimental realizations in optics
imply local non-PT-symmetric potentials \cite{Huang2016}.

The  remaining device is a one-way reflector ($\cal{TR/T}$).
Specifically, if we set $T^l =1, R^l = 1, T^r=-1, R^r = 0$, i.e.
$T^l \neq T^r$ but $\fabsq{T^l}=\fabsq{T^r}=1$,
it can be achieved with a PT-symmetric potential, but it must be non-local, see Table I.
(If we set $T^l = T^r = 1$, local forms of the potential are also possible, as demonstrated in the main text.)
For a nonlocal PT-symmetric potential, $V(x,y) = V(-x,-y)^*$ for all $x,y$.
We assume the form $V (x, y) = \sum_{i=0}^5 \sum_{j=0}^1 v_{ij} x^i y^j$
with $v_{ij} = (-1)^{i+j} v_{ij}^*$, in other words, $v_{ij}$ must be real for $i+j$ even
and purely imaginary for $i+j$ odd.
We also require that $V(-d,y)=V(d,y)=0$ for all $y$ and
follow the same procedure described in previous subsections.
The nonlocal PT-potential found
can  be seen in Fig. \ref{fig_nonlocal_pt} (a),(b) for $k=k_0= 1/d$.
The transmission and reflection coefficients around $k_0$ are depicted in Fig. \ref{fig_nonlocal_pt} (c),(d).
%

\end{document}